\documentclass[11pt,a4paper,numbers,sort&compress]{article}
\usepackage[utf8]{inputenc}
\pdfoutput=1
\usepackage{jheppub}

\usepackage{graphicx} 
\usepackage{amsmath,amsfonts,amssymb}
\usepackage{physics}

\DeclareFontEncoding{LS1}{}{}
\DeclareFontSubstitution{LS1}{stix}{m}{n}
\newcommand*\kay{%
  \text{%
  \fontencoding{LS1}%
  \fontfamily{stixscr}%
  \fontseries{\textmathversion}%
  \fontshape{n}%
  \selectfont\symbol{"6B}}}
\makeatletter
  \newcommand*\textmathversion{\csname textmv@\math@version\endcsname}
  \newcommand*\textmv@normal{m}
  \newcommand*\textmv@bold{b}
\makeatother

\newcommand\fft[2]{\frac{#1}{#2}}
\newcommand\ft[2]{{\textstyle\frac{#1}{#2}}}
\newcommand\nn{\nonumber}

\preprint{LCTP-23-11}

\title{Consistent truncations in higher derivative supergravity}

\author{James T. Liu and Robert J. Saskowski}
\emailAdd{jimliu@umich.edu, rsaskows@umich.edu}

\affiliation{Leinweber Center for Theoretical Physics, Randall Laboratory of Physics\\The University of Michigan, Ann Arbor, MI 48109-1040, USA }

\abstract{We consider the torus reduction of heterotic supergravity in the presence of four-derivative corrections. In particular, the reduction on $T^n$ generically leads to a half-maximal supergravity coupled to $n$ vector multiplets, and we show that it is consistent to truncate out said vector multiplets. This is done by the analysis of both the bosonic equations of motion and the Killing spinor equations.  As an application of the consistent truncation, we examine the four-derivative corrected BPS black string that reduces to a black hole in minimal nine-dimensional supergravity.}
\keywords{}

\date{\today}

\begin{document}
\maketitle
\section{Introduction and summary}
Consistent truncations have played a pivotal role in theoretical physics, ranging from string theory and supergravity to brane-world scenarios. The general principle is that, given a Kaluza-Klein reduction on some compact manifold, one is interested in removing all but a finite number of modes from the infinite Kaluza-Klein tower in such a way as to maintain consistency of the theory, \textit{i.e.}, such that the solutions to the equations of motion of the truncated theory are also solutions of the original theory. The classic example is the Scherk-Schwarz reduction \cite{Scherk:1979zr} wherein the internal space is taken to be a group manifold which becomes the gauge group of the effective lower-dimensional theory. In this case, one can obtain a consistent truncation by restricting to the singlet sector, which enforces consistency via a symmetry principle. More generally, however, in the absence of a manifest symmetry principle, consistent truncations have traditionally been rare and difficult to construct; see for example \cite{Duff:1984hn,Cvetic:2000dm}. In particular, there is no such simple rule for general reductions, and the consistency of a truncation imposes stringent requirements on the field content and couplings of both the higher and lower dimensional theories.

Naturally, there has been much work on non-trivial consistent truncations. In particular, there are many examples of coset reductions, including sphere truncations \cite{deWit:1986oxb,Nastase:1999kf,Lu:1999bc,Cvetic:1999un,Lu:1999bw,Cvetic:1999au,Cvetic:2000dm,Cvetic:2000nc,Lee:2014mla,Nicolai:2011cy,Samtleben:2019zrh,Bonetti:2022gsl} and more general coset reductions \cite{Cvetic:2003jy,House:2005yc,Cassani:2009ck,Cassani:2010na,Bena:2010pr}; in such cases, the massless sector contains charged (non-singlet) fields that one is interested in keeping and care must be taken that these do not source the fields that one wishes to truncate away. There are also examples of reductions wherein one is interested in keeping a finite number of massive modes, such as those on Sasaki-Einstein spaces \cite{Buchel:2006gb,Gauntlett:2009zw,Cassani:2010uw, Liu:2010sa,Gauntlett:2010vu,Skenderis:2010vz,Bah:2010cu,Liu:2010pq,Bah:2010yt} and $T^{1,1}$ \cite{Liu:2011dw}, where one is often interested in keeping massive breathing and squashing modes. Despite the difficulty of finding consistent truncations, there are powerful results. Indeed, one has the conjecture that any warped product AdS$_D\times M_d$ supergravity solution in ten or eleven dimensions has a consistent truncation to a solution of pure gauged supergravity in $D$ dimensions with the same amount of supersymmetry as the original solution \cite{Gauntlett:2007ma}, with additional evidence having been constructed in \cite{Gauntlett:2007sm,OColgain:2011ng,Jeong:2013jfc,Passias:2015gya,Hong:2018amk,MatthewCheung:2019ehr,Larios:2019kbw,Cheung:2022ilc,Couzens:2022aki}. It is also generally believed that truncating to just the massless graviton multiplet is consistent \cite{Pope:1987ad}. Note that all of these results are at the two-derivative level.

An important more recent development has been the use of exceptional field theory \cite{Hohm:2013pua,Hohm:2013vpa,Hohm:2013uia,Hohm:2014fxa,Godazgar:2014nqa}
as a means to construct consistent truncations \cite{Hohm:2014qga}.  The power of exceptional field theory is that consistency is guaranteed by the use of a generalized Scherk-Schwarz reduction.  This has led to many new examples of consistent truncations \cite{Cassani:2016ncu,Malek:2016bpu,Malek:2017njj,Malek:2017cle,Malek:2018zcz,Malek:2019ucd,Cassani:2019vcl,Malek:2020jsa} as well as analysis of the Kaluza-Klein spectra around such truncations 
\cite{Malek:2019eaz,Malek:2020yue,Eloy:2020uix,Varela:2020wty,Cesaro:2020soq,Bobev:2020lsk,Cesaro:2021haf}.  Such developments have in fact put the Gauntlett-Varela conjecture, \cite{Gauntlett:2007ma}, on firm ground.  Nevertheless, despite such enormous progress in the construction of non-trivial consistent truncations, many of the results are currently limited to the leading-order two-derivative theory.

While it seems reasonable that consistency of a truncation at the two-derivative level would imply consistency at higher-derivative order, it is not clear that this necessarily holds.  After all, one possible obstruction could be a higher-derivative coupling between the retained modes and states in the Kaluza-Klein tower.  In the supergravity context, this could potentially show up as additional couplings between the supergravity multiplet and matter multiplets in the spectrum.  To examine this possibility,  we will work specifically in the context of four-derivative heterotic supergravity reduced on a torus.  This is a very standard Kaluza-Klein reduction, and by restricting to zero modes on the torus (\textit {i.e.}, the singlet sector) one is ensured to obtain a consistent truncation.  The bosonic reduction of the four-derivative theory was obtained in \cite{Eloy:2020dko}.

It is important to note that the reduction of heterotic supergravity on $T^n$ leads to a half-maximal supergravity theory in $10-n$ dimensions coupled to $n$ vector multiplets.  The question then arises whether it is consistent to truncate out the vector multiplets, as they naturally arise at the same massless Kaluza-Klein level from the same ten-dimensional fields that give rise to the lower-dimensional supergravity multiplet.  We answer this in the affirmative by explicitly truncating out the bosonic fields in the vector multiplets at the level of their equations of motion as well as their superpartners at the level of the supersymmetry variations.

While we work in general dimensions, the reduction on $T^4$ to six dimensions was considered in \cite{Chang:2021tsj}, which actually performed a truncation to $\mathcal N=(1,0)$ supergravity coupled to one tensor multiplet and four hypermultiplets.  This truncation further reduces the supersymmetry and was indeed shown to be consistent.  On the other hand, our truncation gives $\mathcal N=(1,1)$ supergravity which, in the $\mathcal N=(1,0)$ language corresponds to supergravity coupled to one tensor and two gravitino multiplets.  We show that, while the gravitino multiplet can be consistently truncated, the tensor multiplet cannot be removed at the four-derivative level, even though it can be decoupled from the two-derivative theory.  This is a concrete example of a higher-derivative obstruction to a consistent truncation, even in the relatively simple example of a torus reduction.

\subsection{The torus reduction}

We work with the fields of ten-dimensional heterotic supergravity, $(g_{MN},\psi_M,B_{MN},\lambda,\phi)$, disregarding the heterotic vector multiplets.  Our starting point is the torus reduction of the metric
\begin{equation}
    \dd s_{10}^2=g_{\mu\nu}\dd x^\mu \dd x^\nu+g_{ij}\eta^i\eta^j,\qquad\eta^i=\dd y^i+A_\mu^i\dd x^\mu,
\end{equation}
where $x^\mu$ are coordinates on the base space and $y^i$ are coordinates on the internal space. The two-form $B$ is reduced as
\begin{equation}
    B=\ft12b_{\mu\nu}\dd x^\mu\wedge \dd x^\nu+B_{\mu i}\dd x^\mu\wedge\eta^i+\ft12b_{ij}\eta^i\wedge\eta^j.
\end{equation}
Naturally, reducing the 10D gravity multiplet on an $n$-dimensional torus leads to a half-maximal gravity multiplet coupled to $n$ vector multiplets. By analyzing the bosonic equations of motion, we show that it is consistent to truncate out the vector multiplet, and we write down the reduced Lagrangian.  The resulting bosonic reduction ansatz, $(g_{MN},B_{MN},\phi)\to(g_{\mu\nu},b_{\mu\nu},A_\mu^{(-)\,i},\varphi)$, takes the form
\begin{align}
    &g_{\mu\nu}=g_{\mu\nu},\qquad A_\mu^i=\fft12A_\mu^{(-)\,i},\kern2.9em g_{ij}=\delta_{ij}+\frac{\alpha'}{16}F^{(-)\,i}_{\mu\nu}F^{(-)\,i}_{\mu\nu},\nn\\
    &b_{\mu\nu}=b_{\mu\nu},\qquad B_{\mu i}=-\fft12A_\mu^{(-)\,i},\qquad b_{ij}=0,\nn\\
    &\phi=\varphi.
\label{eq:truncanz1}
\end{align}
Here we have introduced the notation $A^{(\pm)\,i}=A_\mu^i\pm \delta^{ij}B_{\mu i}$, or equivalently $F^{(\pm)\,i}=F^{i}\pm \delta^{ij}G_j$, where $F^i=\dd A^i$ and $G_i=\dd B_i$.  The $A^{(+)\,i}$ are in the vector multiplet and are truncated out along with the scalars $g_{ij}$ and $b_{ij}$.  Note, in particular, the $\mathcal O(\alpha')$ addition to $g_{ij}$ that is required for the truncation to be consistent.

For the fermions, the gravitino $\psi_M$ splits into a lower-dimensional gravitino $\psi_\mu$ along the uncompactified directions and gaugini $\psi_i$ along the compact directions. After an appropriate shift, we show that the truncation of the bosonic sector is consistent with supersymmetry, in the sense that $\delta\tilde\psi_i=0$ where $\tilde\psi_i$ are the $\mathcal O(\alpha')$ corrected gaugini. In particular, if we write the gravitino variation as $\delta\psi_\mu=\mathcal D_\mu\epsilon$, then this redefinition takes the elegant form
\begin{equation}
    \tilde\psi_i=\psi_i-\frac{\alpha'}{4}F^{(-)\,i}_{\mu\nu}\mathcal D_\mu\psi_\nu,
\end{equation}
where $F^{(-)\,i}$ is the combination of field strengths that remains after our truncation, and this is specifically selected out by the gaugino variation. 

\subsection{An \texorpdfstring{$
\mathcal O(\alpha')$}{O(alpha')} corrected black string}

Finally, for illustrative purposes, we look at the four-derivative corrected BPS black string in ten dimensions which reduces to a nine-dimensional black hole.  The leading order black hole solution takes the well-known form \cite{Lu:1995cs}
\begin{align}
    \dd s_9^2&=-\qty(1+\frac{k}{r^6})^{-2}\dd t^2+\dd r^2+r^2\dd\Omega_{7}^2,\nn\\
    A&=\frac{1}{1+\frac{k}{r^6}}\dd t,\nn\\
    e^{\varphi}&=\qty(1+\frac{k}{r^6})^{-1/2}.
\end{align}
We find that the four-derivative corrections to the 10D uplifted metric are then
\begin{align}
    \dd s_{10}^2=&-\qty(1+\frac{k}{r^6})^{-2}\qty(1+\frac{18\alpha' k^2}{r^2(k+r^6)^2})\dd t^2+\dd r^2+r^2\dd\Omega_7^2\nn\\
    &+\qty(1-\frac{18\alpha' k^2}{r^2(k+r^6)^2})\qty(\dd z-\frac{1}{1+\frac{k}{r^6}}\qty(1+\frac{18\alpha' k^2}{r^2(k+r^6)^2})\dd t)^2+\mathcal{O}(\alpha'^2),
\end{align}
while the $B$-field remains unchanged. Similar $\alpha'$-corrected heterotic black holes in lower dimensions were considered in \cite{Natsuume:1994hd,Kats:2006xp,Castro:2007hc,Cvitan:2007pk,Cano:2018brq,Cano:2019ycn,Cano:2021nzo,Ortin:2021win,Cano:2022tmn}. In particular, the truncation places requirements on the components of the metric in the compactified direction, and we find that these are precisely in agreement with the four-derivative corrected black hole solution.

The rest of the paper is organized as follows. In Section \ref{sec:torus}, we review four-derivative heterotic supergravity and discuss the torus reduction. In Section \ref{sec:bosonic}, we verify the consistency of truncating out the vector multiplets by analyzing the bosonic equations of motion, and in Section \ref{sec:fermionic}, we verify the consistency by analysis of the gaugini variations. In Section \ref{sec:example}, we derive the four-derivative corrections to the ten-dimensional BPS black string geometry and compare it with the field redefinitions required in Section \ref{sec:bosonic}. Finally, we conclude in Section \ref{sec:conclusion} and discuss some further truncations.

\section{Heterotic torus reduction}\label{sec:torus}

In this section, we reduce the bosonic fields of four-derivative heterotic supergravity on a torus. Our notation is such that we use $M, N,\ldots$ for curved indices in 10D and $A, B,\ldots$ for rigid indices in 10D, as well as $\mu,\nu,\ldots$ for curved indices along the base space, $\alpha,\beta,\ldots$ for rigid indices along the base space, $i,j,\ldots$ for curved indices along the internal torus, and $a,b,\ldots$ for rigid indices along the internal torus. That is, we split our curved indices as $M\to\{\mu, i\}$ and our rigid indices as $A\to\{\alpha, a\}$. We use $\hat\nabla$ to mean the Levi-Civita connection in 10D, while we use $\nabla$ for the Levi-Civita connection on the base space.

\subsection{Four-derivative heterotic supergravity}
Heterotic supergravity is a ten-dimensional, $\mathcal N=(1,0)$ theory with a single Majorana-Weyl supercharge. The field content is simply the half-maximal gravity multiplet, consisting of the metric $g_{MN}$, the Majorana-Weyl gravitino $\psi_M$, the two-form $B_{MN}$, the Majorana-Weyl dilatino $\lambda$, and the dilaton $\phi$. In the string frame, the ten-dimensional bosonic Lagrangian up to four-derivative corrections takes the form \cite{Bergshoeff:1988nn,Bergshoeff:1989de,Metsaev:1987zx,Chemissany:2007he}
\begin{equation}
    e^{-1}\mathcal L=e^{-2\phi}\left[R+4(\partial_M\phi)^2-\fft1{12}\tilde H_{MNP}^2+\frac{\alpha'}{8}\big(R_{MNAB}(\Omega_+)\big)^2\right]+\mathcal{O}(\alpha'^3),
\label{eq:Lhet}
\end{equation}
where $R$ is the Ricci scalar and we have defined
\begin{equation}
    \tilde H= H-\frac{\alpha'}{4}\omega_{3L}(\Omega_+),
\label{eq:Htilde}
\end{equation}
where $H=\dd B$ is the three-form flux. Note that we have implicitly truncated out the heterotic gauge fields, as they will not play an important role in our discussion. Here we have introduced the torsionful connection
\footnote{Note that the choice of $\Omega_+$ versus $\Omega_-$ is equivalent to a choice of the sign of $H$ in the gravitino variation, and we may always switch conventions by doing a sign flip $B\to -B$. This is just a choice of worldsheet parity.  Our convention is opposite that used in \cite{Bergshoeff:1989de}.}
\begin{equation}
    \Omega_+=\Omega+\fft12\mathcal H,\qquad\mathcal H^{AB}\equiv \tilde H_M{}^{AB}\dd x^M,
\end{equation}
where $\Omega$ is the spin connection, and the corresponding curvature is
\begin{equation}
    R(\Omega_+)=\dd\Omega_++\Omega_+\wedge\Omega_+.
\end{equation}
Such choice of connection is required so that $(\Omega_+,\psi_{MN})$ transforms as an $SO(9,1)$ gauge multiplet \cite{Bergshoeff:1989de}, where $\psi_{MN}=2\nabla_{[M}(\Omega_-)\psi_{N]}$ is the supercovariant gravitino curvature. The Lorentz Chern-Simons form is
\begin{equation}
    \omega_{3L}(\Omega_+)=\Tr\left(\Omega_+\wedge \dd\Omega_++\fft23\Omega_+\wedge\Omega_+\wedge\Omega_+\right),
\label{eq:LCS+}
\end{equation}
and is required by anomaly cancellation. This immediately leads to the Bianchi identity
\begin{equation}
    \dd\tilde H=-\frac{\alpha'}{4}\Tr \qty[R(\Omega_+)\land R(\Omega_+)].
\end{equation}
This is characteristic of the two-group structure.

Note that we can break up the Lagrangian, (\ref{eq:Lhet}), into two- and four-derivative parts
\begin{align}
    e^{-1}\mathcal L_{2\partial}&=e^{-2\phi}\left[R+4(\partial_M\phi)^2-\fft1{12}H_{MNP}^2\right],\nn\\
    e^{-1}\mathcal L_{4\partial}&=\frac{\alpha'}{8}e^{-2\phi}\left[\qty(R_{MNAB}(\Omega_+))^2+\fft13H^{MNP}\omega_{3L\,MNP}(\Omega_+)\right].
\label{eq:Lags}
\end{align}
The bosonic equations of motion are
\begin{align}
    0&=\mathcal E_\phi\equiv R-4\qty(\partial_M\phi)^2+4\hat\Box\phi-\fft1{12}\tilde H_{MNP}^2+\frac{\alpha'}{8}\qty(R_{MNAB}(\Omega_+))^2,\nn\\
    0&=\mathcal E_{g,MN}\equiv R_{MN}+2\hat\nabla_M\hat\nabla_N\phi-\fft14\tilde H_{MAB}\tilde H_N{}^{AB}+\frac{\alpha'}{4}R_{MPAB}(\Omega_+)R_N{}^{PAB}(\Omega_+),\nn\\    0&=\mathcal E_{H,NP}\equiv\hat\nabla^M\qty(e^{-2\phi}\tilde H_{MNP}),
\end{align}
where we have used the dilaton equation $\mathcal E_\phi$ to simplify the Einstein equation $\mathcal E_{g,MN}$. Equivalently, one may make use of the fact that the variation of the action with respect to $\Omega_+$ is proportional to the two-derivative equations of motion \cite{Bergshoeff:1988nn}. These can also be broken up into two- and four-derivative parts, and we write $\mathcal E=\mathcal E^{(0)}+\alpha'\mathcal E^{(1)}$.  Then
\begin{align}
    \mathcal E_\phi^{(0)}&=R-4(\partial_M\phi)^2+4\hat\Box\phi-\fft1{12} H_{MNP}^2,\nn\\
    \mathcal E_{g,MN}^{(0)}&=R_{MN}+2\hat\nabla_M\hat\nabla_N\phi-\fft14 H_{MAB} H_N{}^{AB},\nn\\
    \mathcal E_{H,NP}^{(0)}&=e^{2\phi}\hat\nabla^M\qty(e^{-2\phi} H_{MNP}),
\label{eq:eom2d}
\end{align}
and
\begin{align}
    \mathcal E_\phi^{(1)}&=\fft1{24}H_{MNP}\omega_{3L}^{MNP}(\Omega_+)+\fft18\qty(R_{MNAB}(\Omega_+))^2,\nn\\
    \mathcal E_{g,MN}^{(1)}&=\fft18 H_{MAB} \omega_{3L\,N}{}^{AB}(\Omega_+)+\fft14R_{MPAB}(\Omega_+)R_N{}^{PAB}(\Omega_+),\nn\\
    \mathcal E_{H,NP}^{(1)}&=-\fft14e^{2\phi}\hat\nabla^M\qty(e^{-2\phi} \omega_{3L,MNP}(\Omega_+)).
\label{eq:eom4d}
\end{align}

\subsubsection{Supersymmetry variations}

Although we primarily focus on the reduction of the bosonic fields, the supersymmetry variations of the fermionic fields also need to be considered in order to ensure a consistent truncation.  Up to $\mathcal O(\alpha')$, the supersymmetry transformations of the gravitino and dilatino are
\cite{Strominger:1986uh, Bergshoeff:1989de}%
\footnote{To avoid confusion with $\delta$ denoting $\mathcal O(\alpha')$ corrections, we use $\delta_\epsilon$ for supersymmetry transformations parameterized by a spinor $\epsilon$.}
\begin{align}
    \delta_\epsilon\psi_M&=\nabla_M(\Omega_-)\epsilon=\left(\partial_\mu+\fft14\Omega_{-\,M}{}^{AB}\Gamma_{AB}\right)\epsilon=\left(\nabla_M-\fft18\tilde H_{MNP}\Gamma^{NP}\right)\epsilon,\nn\\
    \delta_\epsilon\lambda&=\left(\Gamma^M\partial_M\phi-\fft1{12}\tilde H_{MNP}\Gamma^{MNP}\right)\epsilon.
\end{align}
The structure of these variations is such that the $\mathcal O(\alpha')$ corrections are entirely contained in the definition of $\tilde H$ given in (\ref{eq:Htilde}).  As above, we can write
\begin{equation}
    \delta_\epsilon \psi_M=\delta_\epsilon\psi_M^{(0)}+\alpha'\delta_\epsilon\psi_M^{(1)},\qquad\delta_\epsilon\lambda=\delta_\epsilon\lambda^{(0)}+\alpha'\delta_\epsilon\lambda^{(1)},
\end{equation}
where
\begin{align}
    \delta_\epsilon\psi_M^{(0)}&=\left(\nabla_M-\fft18H_{MNP}\Gamma^{NP}\right)\epsilon,&\delta_\epsilon\psi_M^{(1)}&=\fft1{32}\omega_{3L,MNP}\Gamma^{NP}\epsilon,\nn\\
    \delta_\epsilon\lambda^{(0)}&=\left(\Gamma^M\partial_M\phi-\fft1{12} H_{MNP}\Gamma^{MNP}\right)\epsilon,&\delta_\epsilon\lambda^{(1)}&=\fft1{48}\omega_{3L,MNP}\Gamma^{MNP}\epsilon.
\label{eq:deltas}
\end{align}

\subsection{Torus reduction}
We perform a standard Kaluza-Klein reduction on an $n$-dimensional torus $T^n$ by taking our metric to be
\begin{equation}
    \dd s_{10}^2=g_{\mu\nu}\dd x^\mu \dd x^\nu+g_{ij}\eta^i\eta^j,\qquad\eta^i=\dd y^i+A_\mu^i\dd x^\mu,
\label{eq:redMet}
\end{equation}
where $x^\mu$ are coordinates on the base space and $y^i$ are coordinates on the internal space. We can introduce a natural zehnbein basis
\begin{equation}
    E^\alpha=e_\mu^\alpha \dd x^\mu,\qquad E^a=e_i^a\eta^i,
\end{equation}
where $e^\alpha$ is a vielbein for $g_{\mu\nu}$ and $e^a$ is a vielbein for $g_{ij}$, so that $\dd s_{10}^2=\eta_{\alpha\beta}E^\alpha E^\beta+\delta_{ab}E^aE^b$.  Then
\begin{equation}
    E=\begin{pmatrix}e^\alpha\\e_i^a\eta^i\end{pmatrix},\qquad \dd E=\begin{pmatrix}\dd e^\alpha\\ \dd e_i^a\wedge\eta^i+e_i^aF^i\end{pmatrix},
\end{equation}
where the abelian field strength is given locally by $F^i=\dd A^i$.  In components, we have
\begin{equation}
    E_M{}^A=\begin{pmatrix}e_\mu^\alpha~&e_i^aA_\mu^i\\0&e_i^a\end{pmatrix},\qquad E_A{}^M=\begin{pmatrix}e_\alpha^\mu~&-e_\alpha^\mu A_\mu^i\\0&e_a^i\end{pmatrix}.
\end{equation}

The torsion-free spin connection can be computed to be
\begin{equation}
    \Omega=\begin{pmatrix}\omega^{\alpha\beta}-\fft12g_{ij}F_{\alpha\beta}^i\eta^j&\fft12e_i^bF_{\mu\alpha}^i\dd x^\mu-\fft12e^i_b\partial_\alpha g_{ij}\eta^j\\-\fft12e_i^aF_{\mu\beta}^i\dd x^\mu+\fft12e^i_a\partial_\beta g_{ij}\eta^j&\fft12\qty(e^{ia}\dd e_i^b-e^{ib}\dd e_i^a)\end{pmatrix},
\label{eq:tfc}
\end{equation}
where $\omega$ is the torsion-free spin connection on the base manifold.

\subsubsection{Inclusion of torsion}

In addition to the metric, the $B$-field is reduced according to
\begin{equation}
    B=\ft12b_{\mu\nu}\dd x^\mu\wedge \dd x^\nu+B_{\mu i}\dd x^\mu\wedge\eta^i+\ft12b_{ij}\eta^i\wedge\eta^j.
\label{eq:redB}
\end{equation}
Computing $H=\dd B$ then gives
\begin{equation}
    H=h+\tilde G_i\wedge\eta^i+\ft12\dd b_{ij}\wedge\eta^i\wedge\eta^j,
\end{equation}
where
\begin{equation}
    h=\dd b-F^i\wedge B_i,\qquad\tilde G_i=G_i-b_{ij}F^j,\qquad G_i=\dd B_i.
\end{equation}
The one-form $\mathcal H^{AB}$ is then
\begin{equation}
    \mathcal H=\begin{pmatrix}h_\mu{}^{\alpha\beta}\dd x^\mu+\tilde G^{\alpha\beta}{}_i\eta^i&e^{ib}\qty(\tilde G_{\mu\alpha i}\dd x^\mu+\partial_\alpha b_{ij}\eta^j)\\-e^{ia}\qty(\tilde G_{\mu\beta i}\dd x^\mu+\partial_\beta b_{ij}\eta^j)&e^{ia}e^{jb}\dd b_{ij}\end{pmatrix}.
\end{equation}
Combining $\mathcal H$ with the torsion-free connection $\Omega$ in (\ref{eq:tfc}) then gives the torsional connection
\begin{align}
    \Omega_+=&\left(\begin{matrix}
        \omega_+^{\alpha\beta}-\fft12\qty(g_{ij}F_{\alpha\beta}^j-\tilde G_{\alpha\beta i})\eta^i\\
        -\fft12e^{ia}\qty(\qty(g_{ij}F_{\mu\beta}^j+\tilde G_{\mu\beta i})dx^\mu-\partial_\beta(g_{ij}-b_{ij})\eta^j)
    \end{matrix}\right.\nn\\
    &\kern12em\left.\begin{matrix}
        \fft12e^{ib}\qty(\qty(g_{ij}F_{\mu\alpha}^j+\tilde G_{\mu\alpha i})dx^\mu-\partial_\alpha\qty(g_{ij}-b_{ij})\eta^j)\\
        \fft12e^{ia}e^{jb}\qty(e_j^c\dd e_i^c-e_i^c\dd e_j^c+\dd b_{ij})
    \end{matrix}\right).
\label{eq:Omega+}
\end{align}
The connection $\Omega_-$ may be obtained by taking $H\to-H$.  The torsionful Riemann tensor can be calculated from $R(\Omega_+)=\dd \Omega_++\Omega_+\wedge\Omega_+$.  The frame components are given in Appendix \ref{app:riemann}.

\subsection{The bosonic reduction at leading order}

Before proceeding with the truncation of the reduced vector multiplets, it is instructive to review the standard Kaluza-Klein reduction of the two-derivative action and equations of motion.  Since the truncation to the zero modes on the torus, (\ref{eq:redMet}) and (\ref{eq:redB}), is guaranteed to be consistent, we can directly reduce the two-derivative Lagrangian, (\ref{eq:Lags}).  This yields the standard Kaluza-Klein result \cite{Maharana:1992my}
\begin{align}
    e^{-1}\mathcal L^{(0)}&=e^{-2\varphi}\Bigl[R(\omega)+4\partial_\mu\varphi^2-\fft1{12}h_{\mu\nu\rho}^2-\fft14\left(g_{ij}F_{\mu\nu}^iF^{\mu\nu\,j}+g^{ij}\tilde G_{\mu\nu\,i}\tilde G^{\mu\nu}_j\right)\nn\\
    &\kern4em-\fft14g^{ij}g^{kl}(\partial_\mu g_{ik}\partial^\mu g_{jl}+\partial_\mu b_{ik}\partial^\mu b_{jl})\Bigr],
\label{eq:redlag2d}
\end{align}
where the reduced dilaton $\varphi$ is given by
\begin{equation}
    \varphi=\phi-\fft14\log\det g_{ij}.
\label{eq:dilshift}
\end{equation}

It is also straightforward to directly reduce the leading order ten-dimensional equations of motion, (\ref{eq:eom2d}).  Making use of some of the reduction expressions in the Appendix, we obtain the reduced two-derivative Einstein equations
\begin{align}
    \mathcal E_{g,\alpha\beta}^{(0)}=&R(\omega)_{\alpha\beta}-\fft12\qty(g_{ij}F_{\alpha\gamma}^iF_{\beta\gamma}^j+g^{ij}\tilde G_{\alpha\gamma\,i}\tilde G_{\beta\gamma\,j})-\fft14h_{\alpha\gamma\delta}h_{\beta\gamma\delta}+2\nabla_\alpha\nabla_\beta\varphi\nonumber\\
    &-\fft14g^{ij}g^{kl}\qty(\partial_\alpha g_{ik}\partial_\beta g_{jl}+\partial_\alpha b_{ik}\partial_\beta b_{jl}),\nn\\
    \mathcal E_{g,\alpha b}^{(0)}=&\fft12e^i_b\left(e^{2\varphi}\nabla_\gamma(e^{-2\varphi}g_{ij}F_{\alpha\gamma}^j)-\fft12h_{\alpha\gamma\delta}\tilde G_{\gamma\delta\,i}-g^{jk}\tilde G_{\alpha\gamma\,j}\partial_\gamma b_{ki}\right),\nn\\
    \mathcal E_{g,ab}^{(0)}=&-\fft12e^i_ae^j_b\biggl(e^{2\varphi}\nabla^\gamma(e^{-2\varphi}\nabla_\gamma) g_{ij}-\fft12(g_{ik}g_{jl}F_{\alpha\beta}^kF_{\alpha\beta}^l-\tilde G_{\alpha\beta\,i}\tilde G_{\alpha\beta\,j})\nn\\
    &\qquad-g^{kl}\qty(\partial_\gamma g_{ik}\partial_\gamma g_{jl}-\partial_\gamma b_{ik}\partial_\gamma b_{jl})\biggr),
\label{eq:eom2dred}
\end{align}
and the reduced $H$-field equations of motion
\begin{align}
    \mathcal E_{H,\alpha\beta}^{(0)}&=e^{2\varphi}\nabla^\gamma\qty(e^{-2\varphi}h_{\alpha\beta\gamma}),\nn\\
    \mathcal E_{H,\alpha b}^{(0)}&=e_i^b\left(e^{2\varphi}\nabla^\gamma(e^{-2\varphi}g^{ij}\tilde G_{\gamma\alpha\,j})+\fft12h_{\alpha\gamma\delta}F_{\gamma\delta}^i\right),\nn\\
    \mathcal E_{H,ab}^{(0)}&=e^{[i}_ae^{j]}_b\left(e^{2\varphi}\nabla^\gamma(e^{-2\varphi}\nabla_\gamma b_{ij})-g_{ik}F_{\gamma\delta}^k\tilde G_{\gamma\delta\,j}+2g^{kl}\partial_\gamma g_{ik}\partial_\gamma b_{jl}\right).
\label{eq:eom2dredH}
\end{align}
Finally, the reduced dilaton equation is
\begin{align}
    \mathcal E_\phi^{(0)}=&R(\omega)-\fft14\qty(g_{ij}F_{\alpha\beta}^iF_{\alpha\beta}^j+g^{ij}\tilde G_{\alpha\beta\,i}\tilde G_{\alpha\beta\,j})-\fft1{12}h_{\alpha\beta\gamma}^2+4\Box\varphi-4\qty(\partial_\alpha\varphi)^2\nn\\
    &-\fft14g^{ij}g^{kl}\qty(\partial_\alpha g_{ik}\partial_\alpha g_{jl}+\partial_\alpha b_{ik}\partial_\alpha b_{jl}).
\label{eq:eom2dredphi}
\end{align}
Since the torus reduction is consistent, these equations can also be directly obtained from the reduced Lagrangian, (\ref{eq:redlag2d}).

\subsection{Supersymmetry variations at leading order}

Along with the leading order bosonic reduction, we can consider the supersymmetry variations of the gravitino and dilatino.  When dimensionally reduced, we have $\{\psi_M,\lambda\}\longrightarrow\{\psi_\mu, \psi_i, \lambda\}$.  As in the case of the lower-dimensional dilaton shift, (\ref{eq:dilshift}), the dilatino also requires a shift of the form
\begin{equation}
    \tilde\lambda=\lambda-\Gamma^i\psi_i.
\end{equation}
With this in mind, the reduction of the lowest-order transformations, (\ref{eq:deltas}), gives
\begin{align}
    \delta_\epsilon\psi_\mu^{(0)}&=\left(\nabla_\mu(\omega_-)+\fft14\qty(g_{ij}F_{\mu\nu}^j-\tilde G_{\mu\nu\,i})\gamma^\nu\Gamma^i-\fft18\qty(2e_i^c\partial_\mu e_j^c+\partial_\mu b_{ij})\Gamma^{ij}\right)\epsilon,\nn\\
    \delta_\epsilon\psi_i^{(0)}&=\left(-\fft18\qty(g_{ij}F_{\mu\nu}^j+\tilde G_{\mu\nu\,i})\gamma^{\mu\nu}-\fft14\partial_\mu\qty(g_{ij}-b_{ij})\gamma^\mu\Gamma^j\right)\epsilon,\nn\\
    \delta_\epsilon\tilde\lambda^{(0)}&=\left(\gamma^\mu\partial_\mu\varphi-\fft1{12}h_{\mu\nu\lambda}\gamma^{\mu\nu\lambda}+\fft18(g_{ij}F_{\mu\nu}^j-\tilde G_{\mu\nu\,i})\gamma^{\mu\nu}\Gamma^i\right)\epsilon.
\label{eq:deltalo}
\end{align}
At this order, the gravitino $\psi_\mu^{(0)}$ and dilatino $\tilde\lambda^{(0)}$, belong in the supergravity multiplet, while the internal components $\psi_i^{(0)}$ fall into vector multiplets.  This allows us to identify the graviphoton and vector multiplet gauge field combinations as
\begin{align}
    F_{\mu\nu}^{a\,(-)}&=e^a_iF_{\mu\nu}^i-e_a^i\tilde G_{\mu\nu\,i},&&(\hbox{graviphoton})\nn\\
    F_{\mu\nu}^{a\,(+)}&=e^a_iF_{\mu\nu}^i+e_a^i\tilde G_{\mu\nu\,i}.&&(\hbox{vector})
\label{eq:gpv}
\end{align}
This will serve as a guide for truncating out the vector multiplets below.

\section{Truncating out the vector multiplets}\label{sec:bosonic}

Reducing the ten-dimensional heterotic action on $T^n$ gives rise to a lower-dimensional half-maximal supergravity coupled to $n$ vector multiplets.  Here we proceed to truncate out the vector multiplets, leading to a pure half-maximal supergravity in lower dimensions.  The truncation of the two-derivative theory is straightforward, and our main intent is to highlight that the truncation remains consistent at the four-derivative level.  We start by considering the two-derivative truncation.

\subsection{The supergravity truncation at leading order}

As indicated in (\ref{eq:deltalo}) and (\ref{eq:gpv}), the bosonic fields in the vector multiplet consist of the vectors $F_{\mu\nu}^{a\,(+)}$ along with their scalar superpartners $g_{ij}-b_{ij}$.  This suggests that, at least at leading order, we can truncate out the vector multiplets by taking
\begin{equation}
    g_{ij}=\delta_{ij},\qquad b_{ij}=0,\qquad G_{\mu\nu\,i}=-F_{\mu\nu}^i.
\end{equation}
(Note that, with $g_{ij}=\delta_{ij}$, the internal indices $i,j,\ldots$ are raised and lowered using $\delta_{ij}$.)  However, as an intermediate step, it is instructive to truncate the scalars first before considering the gauge fields.  Thus we let
\begin{equation}
    g_{ij}=\delta_{ij},\qquad b_{ij}=0,\qquad F_{\mu\nu}^{(\pm)\,i}=F_{\mu\nu}^i\pm G_{\mu\nu\,i}.
\end{equation}
In this case, the two-derivative equations of motion, (\ref{eq:eom2dred}), (\ref{eq:eom2dredH}) and (\ref{eq:eom2dredphi}), take the form
\begin{align}
    \mathcal E_{g,\alpha\beta}^{(0)}&=R(\omega)_{\alpha\beta}-\fft14\qty(F_{\alpha\gamma}^{(+)\,i}F_{\beta\gamma}^{(+)\,i}+F_{\alpha\gamma}^{(-)\,i}F_{\beta\gamma}^{(-)\,i})-\fft14h_{\alpha\gamma\delta}h_{\beta\gamma\delta}+2\nabla_\alpha\nabla_\beta\varphi,\nn\\
    \mathcal E_{g,\alpha i}^{(0)}&=-\fft14\left(e^{2\varphi}\nabla_\gamma(e^{-2\varphi}F_{\gamma\alpha}^{(+)\,i})+\fft12h_{\alpha\gamma\delta} F_{\gamma\delta}^{(+)\,i}\right)-\fft14\left(e^{2\varphi}\nabla_\gamma(e^{-2\varphi}F_{\gamma\alpha}^{(-)\,i})-\fft12h_{\alpha\gamma\delta} F_{\gamma\delta}^{(-)\,i}\right),\nn\\
    \mathcal E_{g,ij}^{(0)}&=\fft18\qty(F_{\alpha\beta}^{(+)\,i}F_{\alpha\beta}^{(-)\,j}+F_{\alpha\beta}^{(-)\,i}F_{\alpha\beta}^{(+)\,j}),\nn\\
    \mathcal E_\phi^{(0)}&=R(\omega)-\fft18\qty(F_{\alpha\beta}^{(+)\,i}F_{\alpha\beta}^{(+)\,i}+F_{\alpha\beta}^{(-)\,i}F_{\alpha\beta}^{(-)\,i})-\fft1{12}h_{\alpha\beta\gamma}^2+4\Box\varphi-4(\partial_\alpha\varphi)^2,\nn\\
    \mathcal E_{H,\alpha\beta}^{(0)}&=e^{2\varphi}\nabla^\gamma\qty(e^{-2\varphi}h_{\alpha\beta\gamma}),\nn\\
    \mathcal E_{H,\alpha i}^{(0)}&=\fft12\left(e^{2\varphi}\nabla^\gamma(e^{-2\varphi}F_{\gamma\alpha}^{(+)\,i})+\fft12h_{\alpha\gamma\delta}F_{\gamma\delta}^{(+)\,i}\right)-\fft12\left(e^{2\varphi}\nabla^\gamma(e^{-2\varphi}F_{\gamma\alpha}^{(-)\,i})-\fft12h_{\alpha\gamma\delta}F_{\gamma\delta}^{(-)\,i}\right),\nn\\
    \mathcal E_{H,ij}^{(0)}&=\fft14\qty(F_{\alpha\beta}^{(+)\,i}F_{\alpha\beta}^{(-)\,j}-F_{\alpha\beta}^{(-)\,i}F_{\alpha\beta}^{(+)\,j}).
\end{align}

At the bosonic level, we can proceed in two ways, by either setting $F^{(+)}=0$ or $F^{(-)}=0$.  The former case will truncate out the gauge fields in the vector multiplet, while the latter will remove the graviphotons, leading to a consistent but non-supersymmetric truncation.  Note, in particular, that the two-derivative bosonic Lagrangian, (\ref{eq:Lags}), is invariant under $B\to-B$.  This is what underlies the symmetry between $F^{(+)}$ and $F^{(-)}$ at the leading order.

We are, of course, mainly interested in a supersymmetric consistent truncation.  Thus we proceed by setting $F^{(+)}=0$.  Specifically, we take
\begin{equation}
    g_{ij}=\delta_{ij},\qquad b_{ij}=0,\qquad A_\mu^i=\fft12A_\mu^{(-)\,i},\qquad B_{\mu\,i}=-\fft12A_\mu^{(-)\,i}.
\label{eq:red2d}
\end{equation}
Doing so then yields the two-derivative equations of motion
\begin{align}
    \mathcal E_{g,\alpha\beta}^{(0)}&=R(\omega)_{\alpha\beta}-\fft14F_{\alpha\gamma}^{(-)\,i}F_{\beta\gamma}^{(-)\,i}-\fft14h_{\alpha\gamma\delta}h_{\beta\gamma\delta}+2\nabla_\alpha\nabla_\beta\varphi,\nn\\
    \mathcal E_{g,\alpha i}^{(0)}&=-\fft14\left(e^{2\varphi}\nabla_\gamma\qty(e^{-2\varphi}F_{\gamma\alpha}^{(-)\,i})-\fft12h_{\alpha\gamma\delta} F_{\gamma\delta}^{(-)\,i}\right),\nn\\
    \mathcal E_{g,ij}^{(0)}&=0,\nn\\
    \mathcal E_\phi^{(0)}&=R(\omega)-\fft18F_{\alpha\beta}^{(-)\,i}F_{\alpha\beta}^{(-)\,i}-\fft1{12}h_{\alpha\beta\gamma}^2+4\Box\varphi-4(\partial_\alpha\varphi)^2,\nn\\
    \mathcal E_{H,\alpha\beta}^{(0)}&=e^{2\varphi}\nabla^\gamma\qty(e^{-2\varphi}h_{\alpha\beta\gamma}),\nn\\
    \mathcal E_{H,\alpha i}^{(0)}&=-\fft12\left(e^{2\varphi}\nabla^\gamma\qty(e^{-2\varphi}F_{\gamma\alpha}^{(-)\,i})-\fft12h_{\alpha\gamma\delta}F_{\gamma\delta}^{(-)\,i}\right),\nn\\
    \mathcal E_{H,ij}^{(0)}&=0.
\label{eq:2dtrunc}
\end{align}
Note, in particular, that the internal Einstein and $H$ equations are trivial, and that the mixed Einstein and $H$ equations are consistent with each other.  This set of equations can be derived from the reduced Lagrangian
\begin{equation}
    e^{-1}\mathcal L=e^{-2\varphi}\left(R+4\qty(\partial\varphi)^2-\fft1{12}h_{\mu\nu\rho}^2-\fft18\qty(F_{\mu\nu}^{(-)\,i})^2\right),
\end{equation}
where the $h$ Bianchi identity is given by
\begin{equation}
    h=\dd b+\fft14F^{(-)\,i}\wedge A^{(-)\,i}\qquad\Rightarrow\qquad \dd h=\fft14F^{(-)\,i}\wedge F^{(-)\,i}.
\label{eq:hbian}
\end{equation}
This can equally well be obtained by directly substituting the truncation ansatz, (\ref{eq:red2d}), into the two-derivative Lagrangian (\ref{eq:Lags}).

\subsection{The supergravity truncation at \texorpdfstring{$\mathcal O(\alpha')$}{O(alpha prime)}}

We now wish to extend the truncation of the vector multiplets to the four-derivative level.  Working to $\mathcal O(\alpha')$, the supergravity truncation, (\ref{eq:red2d}), is expected to receive corrections.  With a slight abuse of notation, we thus let
\begin{align}
    &g_{\mu\nu}=g_{\mu\nu}+\alpha'\delta g_{\mu\nu},\qquad b_{\mu\nu}=b_{\mu\nu}+\alpha'\delta b_{\mu\nu},\qquad \varphi=\varphi+\alpha'\delta\varphi,\nn\\
    &A_\mu^i=\fft12A_\mu^{(-)\,i}+\alpha'\delta A_\mu^i,\qquad B_{\mu i}=-\fft12A_\mu^{(-)\,i}+\alpha'\delta B_{\mu i},\nn\\
    &g_{ij}=\delta_{ij}+\alpha'\delta g_{ij},\qquad b_{ij}=0+\alpha'\delta b_{ij}.
\label{eq:corr}
\end{align}
The equations of motion to $\mathcal O(\alpha')$ then take the form
\begin{equation}
    \mathcal E=\mathcal E^{(0)}+\alpha'\qty(\delta\mathcal E^{(0)}+\mathcal E^{(1)}).
\end{equation}
Here $\delta\mathcal E^{(0)}$ arises from inserting the corrected fields into the two-derivative equations and $\mathcal E^{(1)}$ can be obtained from inserting the leading order fields into the four-derivative equations.

Extending the leading order equations of motion, (\ref{eq:2dtrunc}), to the next order, we see that the necessary conditions for maintaining a consistent truncation are
\begin{equation}
    \delta\mathcal E_{g,ij}^{(0)}+\mathcal E_{g,ij}^{(1)}=0,\qquad \delta\mathcal E_{H,ij}^{(0)}+\mathcal E_{H,ij}^{(1)}=0,
\label{eq:seoms}
\end{equation}
to ensure truncation of the scalars, and
\begin{equation}
    \delta\mathcal E_{g,\alpha i}^{(0)}+\mathcal E_{g,\alpha i}^{(1)}=\fft12\qty(\delta\mathcal E_{H,\alpha i}^{(0)}+\mathcal E_{H,\alpha i}^{(1)}).
\label{eq:max2}
\end{equation}
to ensure truncation of the vector multiplet gauge fields.  Solving these conditions will provide constraints on the correction terms in (\ref{eq:corr}).

To calculate $\mathcal E^{(1)}$, we only need to work with the leading order truncation.  This simplifies various objects needed in the calculation.  In particular, the torsionful spin connection reduces to
\begin{equation}
    \Omega_+=\begin{pmatrix}\omega_+^{\alpha\beta}-\fft12F_{\alpha\beta}^{(-)\,i}\eta^i&0\\0&0\end{pmatrix}.
\end{equation}
This gives the torsionful Riemann tensor
\begin{align}
    R_{\gamma\delta}{}^{\alpha\beta}(\Omega_+)&=R_{\gamma\delta}{}^{\alpha\beta}(\omega_+)-\fft14F_{\gamma\delta}^{(-)\,i}F_{\alpha\beta}^{(-)\,i},\nn\\
    R_{\gamma d}{}^{\alpha\beta}(\Omega_+)&=-\fft12\delta^i_d\nabla_\gamma^{(+)}F_{\alpha\beta}^{(-)\,i},\nn\\
    R_{cd}{}^{\alpha\beta}(\Omega_+)&=\fft12\delta_c^{[i}\delta_d^{j]}F_{\alpha\gamma}^{(-)\,i}F_{\gamma\beta}^{(-)\,j},
\end{align}
and Lorentz Chern-Simons form
\begin{align}
    \omega_{3L,\alpha\beta\gamma}(\Omega_+)&=\omega_{3L,\alpha\beta\gamma}(\omega_+),\nn\\
    \omega_{3L,\alpha\beta c}(\Omega_+)&=\delta^i_c\qty(R_{\alpha\beta}{}^{\gamma\delta}(\omega_+)F_{\gamma\delta}^{(-)\,i}-\fft18F_{\alpha\beta}^{(-)\,j}F_{\gamma\delta}^{(-)\,j}F_{\gamma\delta}^{(-)\,i}),\nn\\
    \omega_{3L,\alpha bc}(\Omega_+)&=\delta^{[i}_b\delta^{j]}_c\qty(\fft12F_{\gamma\delta}^{(-)\,i}\nabla_\alpha^{(+)}F_{\gamma\delta}^{(-)\,j}),\nn\\
    \omega_{3L,abc}(\Omega_+)&=\delta_a^{[i}\delta_b^j\delta_c^{k]}\qty(-\fft12F_{\alpha\beta}^{(-)\,i}F_{\beta\gamma}^{(-)\,j}F_{\gamma\alpha}^{(-)\,k}).
\label{eq:LCSf}
\end{align}
Note that we have dropped an exact term from $\omega_{3L}(\Omega_+)$, which is implicitly absorbed into a field redefinition of $B$.  For details, see Appendix~\ref{sec:CSterm}

\subsubsection{Truncating the internal Einstein equation}
We first check the scalar equations of motion, (\ref{eq:seoms}), corresponding to the internal Einstein equation.  Starting with the $\mathcal E_{g,MN}^{(1)}$ from (\ref{eq:eom4d}), we find
\begin{align}
    \mathcal E_{g,ij}^{(1)}&=\fft1{16}\Bigl[-R_{\alpha\beta\gamma\delta}(\omega_+)F_{\alpha\beta}^{(-)\,(i}F_{\gamma\delta}^{(-)\,j)}+\nabla_\gamma^{(+)}F_{\alpha\beta}^{(-)\,i}\nabla_\gamma^{(+)}F_{\alpha\beta}^{(-)\,j}+\fft18F_{\alpha\beta}^{(-)\,i}F_{\alpha\beta}^{(-)\,k}F_{\gamma\delta}^{(-)\,j}F_{\gamma\delta}^{(-)\,k}\nn\\
    &\qquad+\fft12F_{\alpha\beta}^{(-)\,i}F_{\beta\gamma}^{(-)\,j}F_{\gamma\delta}^{(-)\,k}F_{\delta\alpha}^{(-)\,k}-\fft12F_{\alpha\beta}^{(-)\,i}F_{\beta\gamma}^{(-)\,k}F_{\gamma\delta}^{(-)\,j}F_{\delta\alpha}^{(-)\,k}\Bigr].
\end{align}
Since this is non-zero, it would have to cancel against a similar expression in $\delta\mathcal E_{g,ij}^{(0)}$.  To find this correction, we need to start from the full expression for $\mathcal E_{g,ij}^{(0)}$ in (\ref{eq:eom2dred}).  To first order, we find
\begin{equation}
    \delta\mathcal E_{g,ij}^{(0)}=-\fft12e^{2\varphi}\partial_\gamma\qty(e^{-2\varphi}\partial_\gamma\delta g_{ij})+\fft18F_{\alpha\beta}^{(-)\,k}F_{\alpha\beta}^{(-)\,j}\delta g_{ik}+\fft14F_{\alpha\beta}^{(-)\,i}\qty(\delta F_{\alpha\beta}^j+\delta G_{\alpha\beta\,j}-\fft12F_{\alpha\beta}^{(-)\,k}\delta b_{jk}),
\label{eq:dEgij}
\end{equation}
where symmetry of $(ij)$ is implied.  Our main focus is on $\delta g_{ij}$ and $\delta b_{ij}$.  Since these carry $i$ and $j$ indices, and since we want them to be two-derivative terms, we expect them to be built out of bilinears in the field strengths, $F_{\alpha\beta}^{(-)\,i}$.  We will confirm below that an appropriate choice is to take
\begin{equation}
    \delta g_{ij}=\fft1{16}F_{\alpha\beta}^{(-)\,i}F_{\alpha\beta}^{(-)\,j},\qquad\delta b_{ij}=0.
\label{eq:deltagb}
\end{equation}
Note that there is no obvious antisymmetric choice for $\delta b_{ij}$ so the only natural result is to set it to zero.

After some manipulation, we find
\begin{align}
    -\fft12e^{2\varphi}\partial_\gamma\qty(e^{-2\varphi}\partial_\gamma\delta g_{ij})&=-\fft1{16}\Bigl[-R_{\alpha\beta\gamma\delta}F_{\alpha\beta}^{(-)\,i}F_{\gamma\delta}^{(-)\,j}+\nabla_\gamma F_{\alpha\beta}^{(-)\,i}\nabla_\gamma F_{\alpha\beta}^{(-)\,j}\nn\\
    &\qquad+2F_{\alpha\gamma}^{(-)\,i}F_{\beta\gamma}^{(-)\,j}\qty(R_{\alpha\beta}+2\nabla_\alpha\nabla_\beta\varphi)+2F_{\alpha\beta}^{(-)\,i}\nabla_\alpha\qty(e^{2\varphi}\nabla_\gamma(e^{-2\varphi}F_{\gamma\beta}^{(-)\,j}))\Bigl].
\end{align}
The terms in parentheses in the second line are almost the leading order equations of motion, (\ref{eq:2dtrunc}), but are missing a few terms.  By adding and subtracting, we can arrive at
\begin{align}
    -\fft12e^{2\varphi}\partial_\gamma\qty(e^{-2\varphi}\partial_\gamma\delta g_{ij})&=-\fft1{16}\Bigl[-R_{\alpha\beta\gamma\delta}F_{\alpha\beta}^{(-)\,i}F_{\gamma\delta}^{(-)\,j}+\nabla_\gamma F_{\alpha\beta}^{(-)\,i}\nabla_\gamma F_{\alpha\beta}^{(-)\,j}+\fft12F_{\alpha\beta}^{(-)\,i}F_{\beta\gamma}^{(-)\,j}F_{\gamma\delta}^{(-)\,k}F_{\delta\alpha}^{(-)\,k}\nn\\
    &\qquad+\fft12h_{\alpha\delta\epsilon}h_{\beta\delta\epsilon}F_{\alpha\gamma}^{(-)\,i}F_{\beta\gamma}^{(-)\,j}+F_{\alpha\beta}^{(-)\,i}F_{\gamma\delta}^{(-)\,j}\nabla_\alpha h_{\beta\gamma\delta}+h_{\beta\gamma\delta}F_{\alpha\beta}^{(-)\,i}\nabla_\alpha F_{\gamma\delta}^{(-)\,j}\nn\\
    &\qquad+2F_{\alpha\gamma}^{(-)\,i}F_{\beta\gamma}^{(-)\,j}\mathcal E_{\alpha\beta}^{(0)}+2F_{\alpha\beta}^{(-)\,i}\nabla_\alpha\mathcal E_\beta^{(0)\,j}\Bigr],
\end{align}
where we have normalized the graviphoton equation of motion according to $\mathcal E_\alpha^{(0)\,i}=-4\mathcal E_{g,\alpha i}^{(0)}=-2\mathcal E_{H,\alpha i}^{(0)}$.  We can rewrite the torsion-free Riemann and covariant derivatives in terms of their torsionful versions.  The result is
\begin{align}
    -\fft12e^{2\varphi}\partial_\gamma\qty(e^{-2\varphi}\partial_\gamma\delta g_{ij})&=-\fft1{16}\Bigl[-R_{\alpha\beta\gamma\delta}(\omega_+)F_{\alpha\beta}^{(-)\,i}F_{\gamma\delta}^{(-)\,j}+\nabla_\gamma^{(+)}F_{\alpha\beta}^{(-)\,i}\nabla_\gamma^{(+)}F_{\alpha\beta}^{(-)\,j}\nn\\
    &\qquad+\fft12F_{\alpha\beta}^{(-)\,i}F_{\beta\gamma}^{(-)\,j}F_{\gamma\delta}^{(-)\,k}F_{\delta\alpha}^{(-)\,k}+2F_{\alpha\beta}^{(-)\,i}F_{\gamma\delta}^{(-)\,j}\nabla_\alpha h_{\beta\gamma\delta}\nn\\
    &\qquad+2F_{\alpha\gamma}^{(-)\,i}F_{\beta\gamma}^{(-)\,j}\mathcal E_{\alpha\beta}^{(0)}+2F_{\alpha\beta}^{(-)\,i}\nabla_\alpha\mathcal E_\beta^{(0)\,j}\Bigr],
\end{align}
Taking into account the implicit symmetrization of $(ij)$, the term involving $\nabla_\alpha h_{\beta\gamma\delta}$ can be simplified using the Bianchi identity (\ref{eq:hbian}).  The result is then
\begin{align}
    -\fft12e^{2\varphi}\partial_\gamma\qty(e^{-2\varphi}\partial_\gamma\delta g_{ij})=&-\fft1{16}\Bigl[-R_{\alpha\beta\gamma\delta}(\omega_+)F_{\alpha\beta}^{(-)\,i}F_{\gamma\delta}^{(-)\,j}+\nabla_\gamma^{(+)}F_{\alpha\beta}^{(-)\,i}\nabla_\gamma^{(+)}F_{\alpha\beta}^{(-)\,j}\nn\\
    &\qquad+\fft14F_{\alpha\beta}^{(-)\,i}F_{\alpha\beta}^{(-)\,k}F_{\gamma\delta}^{(-)\,j}F_{\gamma\delta}^{(-)\,k}+\fft12F_{\alpha\beta}^{(-)\,i}F_{\beta\gamma}^{(-)\,j}F_{\gamma\delta}^{(-)\,k}F_{\delta\alpha}^{(-)\,k}\nn\\
    &\qquad-\fft12F_{\alpha\beta}^{(-)\,i}F_{\beta\gamma}^{(-)\,k}F_{\gamma\delta}^{(-)\,j}F_{\delta\alpha}^{(-)\,k}+2F_{\alpha\gamma}^{(-)\,i}F_{\beta\gamma}^{(-)\,j}\mathcal E_{\alpha\beta}^{(0)}+2F_{\alpha\beta}^{(-)\,i}\nabla_\alpha\mathcal E_\beta^{(0)\,j}\Bigl],
\end{align}
Inserting this into (\ref{eq:dEgij}) and taking into account the second term in (\ref{eq:dEgij}) as well, we find
\begin{align}
    \delta\mathcal E_{g,ij}^{(0)}&=-\fft1{16}\Bigl[-R_{\alpha\beta\gamma\delta}(\omega_+)F_{\alpha\beta}^{(-)\,i}F_{\gamma\delta}^{(-)\,j}+\nabla_\gamma^{(+)}F_{\alpha\beta}^{(-)\,i}\nabla_\gamma^{(+)}F_{\alpha\beta}^{(-)\,j}+\fft18F_{\alpha\beta}^{(-)\,i}F_{\alpha\beta}^{(-)\,k}F_{\gamma\delta}^{(-)\,j}F_{\gamma\delta}^{(-)\,k}\nn\\
    &\qquad+\fft12F_{\alpha\beta}^{(-)\,i}F_{\beta\gamma}^{(-)\,j}F_{\gamma\delta}^{(-)\,k}F_{\delta\alpha}^{(-)\,k}-\fft12F_{\alpha\beta}^{(-)\,i}F_{\beta\gamma}^{(-)\,k}F_{\gamma\delta}^{(-)\,j}F_{\delta\alpha}^{(-)\,k}\nn\\
    &\qquad+2F_{\alpha\gamma}^{(-)\,i}F_{\beta\gamma}^{(-)\,j}\mathcal E_{\alpha\beta}^{(0)}+2F_{\alpha\beta}^{(-)\,i}\nabla_\alpha\mathcal E_\beta^{(0)\,j}-4F_{\alpha\beta}^{(-)\,i}\qty(\delta F_{\alpha\beta}^j+\delta G_{\alpha\beta\,j})\Bigr],
\end{align}
As a result, we are left with
\begin{equation}
     \delta\mathcal E_{g,ij}^{(0)}+\mathcal E_{g,ij}^{(1)}=-\fft18\left[F_{\alpha\gamma}^{(-)\,i}F_{\beta\gamma}^{(-)\,j}\mathcal E_{\alpha\beta}^{(0)}+F_{\alpha\beta}^{(-)\,i}\nabla_\alpha\mathcal E_\beta^{(0)\,j}-2F_{\alpha\beta}^{(-)\,i}\qty(\delta F_{\alpha\beta}^j+\delta G_{\alpha\beta\,j})\right],
\end{equation}
which vanishes by the leading order equations of motion, provided
\begin{equation}
    \delta F_{\alpha\beta}^j+\delta G_{\alpha\beta\,j}=0.
\label{eq:deltaFG}
\end{equation}

\subsubsection{Truncating the internal \texorpdfstring{$H$}{H} equation}

We now turn to the internal components of the $H$ equation of motion, (\ref{eq:seoms}).  For $\mathcal E_{H,ij}^{(1)}$, we find
\begin{equation}
    \mathcal E_{H,ij}^{(1)}=-\fft14\Bigg[\fft12e^{2\phi}\nabla_\gamma\qty(e^{-2\phi}F_{\alpha\beta}^{(-)\,i}\nabla_\gamma F_{\alpha\beta}^{(-)\,j})+\fft12e^{2\phi}\nabla_\gamma\qty(e^{-2\phi}h_{\gamma\alpha\beta}F_{\alpha\delta}^{(-)\,i}F_{\beta\delta}^{(-)\,j})-\fft12\nabla_\gamma h_{\delta\alpha\beta}F_{\gamma\delta}^{(-)\,i}F_{\alpha\beta}^{(-)\,j}\Bigg].
\label{eq:cEHij1}
\end{equation}
Note that here we are implicitly assuming antisymmetry on $[ij]$.  This antisymmetry will be very useful in making many terms disappear.  Along with $\mathcal E_{H,ij}^{(1)}$, we also have
\begin{equation}
    \delta\mathcal E_{H,ij}^{(0)}=e^{2\varphi}\nabla^\gamma\qty(e^{-2\varphi}\nabla_\gamma\delta b_{ij})+\fft14F_{\alpha\beta}^{(-)\,j}F_{\alpha\beta}^{(-)\,k}\delta g_{ik}+\fft12F_{\alpha\beta}^{(-)\,j}\qty(\delta F_{\alpha\beta}^i+\delta G_{\alpha\beta\,i}-\fft12F_{\alpha\beta}^{(-)\,k}\delta b_{ik}).
\end{equation}
If we take (\ref{eq:deltagb}) for the corrections $\delta g_{ij}$ and $\delta b_{ij}$, along with (\ref{eq:deltaFG}), we see that this actually vanishes, namely $\delta\mathcal E_{H,ij}^{(0)}=0$.  Thus, to be consistent, we then need to have $\mathcal E_{H,ij}^{(1)}$ vanishing as well.  To see that this is indeed the case, we can manipulate (\ref{eq:cEHij1}) by expanding out the $\nabla_\gamma$ derivative in the first two terms, while making use of antisymmetry on $[ij]$
\begin{align}
    \mathcal E_{H,ij}^{(1)}&=-\fft14\Bigg[\fft12e^{2\phi}F_{\alpha\beta}^{(-)\,i}\nabla_\gamma\qty(e^{-2\phi}\nabla_\gamma F_{\alpha\beta}^{(-)\,j})+\fft12e^{2\phi}\nabla_\gamma\qty(e^{-2\phi}h_{\gamma\alpha\beta})F_{\alpha\delta}^{(-)\,i}F_{\beta\delta}^{(-)\,j}\nn\\
    &\qquad+h_{\gamma\alpha\beta}\nabla_\gamma\qty(F_{\alpha\delta}^{(-)\,i}F_{\beta\delta}^{(-)\,j})-\fft12\nabla_\gamma h_{\delta\alpha\beta}F_{\gamma\delta}^{(-)\,i}F_{\alpha\beta}^{(-)\,j}\Bigg].
\end{align}
We can rewrite the first term using the Bianchi identity $\dd F^{(-)\,j}=0$
\begin{equation}
    \fft12e^{2\phi}F_{\alpha\beta}^{(-)\,i}\nabla_\gamma\qty(e^{-2\phi}\nabla_\gamma F_{\alpha\beta}^{(-)\,j})=e^{2\phi}F_{\alpha\beta}^{(-)\,i}\nabla_\gamma\qty(e^{-2\phi}\nabla_\alpha F_{\gamma\beta}^{(-)\,j}).
\end{equation}
After moving $\nabla_\alpha$ past the dilaton factor and commuting it past the $\nabla_\gamma$, we end up with part of the graviphoton equation of motion, $\mathcal E_\beta^{(0)\,j}$.  Collecting terms and simplifying then gives
\begin{equation}
    \mathcal E_{H,ij}^{(1)}=-\fft14\left[F_{\alpha\beta}^{(-)\,i}\nabla_\alpha\mathcal E_\beta^{(0)\,j}+\fft12\mathcal E_{H,\alpha\beta}^{(0)}F_{\alpha\delta}^{(-)\,i}F_{\beta\delta}^{(-)\,j}\right].
\end{equation}
This now vanishes by the lowest order equations of motion.

\subsubsection{Compatibility of the Maxwell equations}

The final expression to verify is (\ref{eq:max2}), namely the consistency of the two Maxwell equations. The shifts of the two-derivative equations of motion are straightforwardly found to be
\begin{align}
    \delta\mathcal E^{(0)}_{g,\alpha a}&=\delta e^i_a\qty[\frac{1}{2}e^{2\varphi}\nabla^\gamma\qty(e^{-2\varphi}g_{ij}F^{j}_{\alpha\gamma})-\frac{1}{4}h_{\alpha\gamma\delta}G^{\gamma\delta}_i]+\frac{1}{2}e^i_ae^{2\varphi}\nabla^\gamma\qty(e^{-2\varphi}\delta g_{ij} F^j_{\alpha\gamma}),\nonumber\\
    \delta\mathcal E^{(0)}_{H,\alpha a}&=\delta e^a_i\qty[e^{2\varphi}\nabla^\gamma\qty(e^{-2\varphi}g^{ij}G_{j\alpha\gamma})+\frac{1}{2}h_{\alpha\gamma\delta}F^{i\gamma\delta}]+e^a_ie^{2\varphi}\nabla^\gamma\qty(e^{-2\varphi}\delta g^{ij}G_{j\alpha\gamma}),
\end{align}
so that the difference, after imposing our truncation, is simply
\begin{align}
    \delta\mathcal E^{(0)}_{H,\alpha a}-2\delta\mathcal E^{(0)}_{g,\alpha a}=&\delta e^a_i \delta^i_a\mathcal E^{(0)}_{H,\alpha a}-2\delta e^i_a\delta^a_i\mathcal E^{(0)}_{g,\alpha a}+\frac{1}{32}e_i^a h_{\alpha\gamma\epsilon}F^{(-)\,j}_{\gamma\epsilon}F^{(-)\,i}_{\beta\delta}F^{(-)\,j}_{\beta\delta}\nn\\
    &+\frac{1}{8}\delta_i^a F^{(-)\,j}_{\gamma\alpha}F^{(-)\,(i}_{\beta\delta}\nabla_\gamma F^{(-)\,j)}_{\beta\delta}.
\end{align}
The four-derivative parts of the equations of motion are simply
\begin{align}
    \mathcal E^{(1)}_{g,\alpha a}=&\delta_i^a\bigg[\frac{1}{16} R_{\beta\gamma} ^{\ \ \ \delta\epsilon}(\omega_+)h_{\alpha\beta\gamma}F_{\delta\epsilon}^{(-)\,i}+\frac{1}{8}R_{\alpha\beta}^{\ \ \gamma\delta}(\omega_+)\nabla_\beta^{(+)}F^{(-)\,i}_{\gamma\delta}-\frac{1}{32}F^{(-)\,i}_{\beta\gamma}\omega_{3L,\alpha\beta\gamma}(\omega_+)\nn\\
    &\ \ \ -\frac{1}{128} h_{\alpha\beta\gamma}F^{(-)\,j}_{\beta\gamma}F^{(-)\,j}_{\delta\epsilon}F^{(-)\,i}_{\delta\epsilon}-\frac{1}{16}F^{(-)\,[i}_{\gamma\epsilon}F^{(-)\,j]}_{\epsilon\delta}\nabla_\alpha^{(+)}F_{\gamma\delta}^{(-)\,j}\bigg],
\end{align}
and
\begin{align}
    \mathcal E^{(1)}_{H,\alpha a}=-\frac{1}{4}\delta_i^a e^{2\varphi}\nabla^\beta\qty[e^{-2\varphi}\qty(R_{\beta\alpha}^{\ \ \ \gamma\delta}(\omega_+)F_{\gamma\delta}^{(-)\,i}-\frac{1}{8}F^{(-)\,j}_{\beta\alpha}F^{(-)\,j}_{\gamma\delta}F^{(-)\,i}_{\gamma\delta})]-\frac{1}{16}\delta_i^aF^{(-)\,i}_{\beta\gamma}\omega_{3L,\alpha\beta\gamma}(\omega_+).
\end{align}
By use of the torsion-free differential Bianchi identity $\nabla_{[\alpha} R_{\beta\gamma]\delta\epsilon}=0$, we have that \begin{equation}
    \nabla^\beta R_{\alpha\beta\gamma\delta}=-2\nabla_{[\gamma}R_{\delta]\alpha},
\end{equation} 
and so, after appropriate substitution of equations of motion and use of the $h$ Bianchi identity \eqref{eq:hbian}, we find that
\begin{align}
e^{2\varphi}\nabla^\beta\qty(e^{-2\varphi}R_{\beta\alpha}^{\ \ \ \gamma\delta}(\omega_+)F_{\gamma\delta}^{(-)\,i})=&F^{(-)\,i}_{\gamma\delta}\Bigg[-\frac{1}{4}\nabla_\gamma\qty(F^{(-)\,j}_{\delta\epsilon}F^{(-)\,j}_{\epsilon\alpha})-\frac{1}{4}\nabla_\gamma\qty(h_{\delta\beta\epsilon}h_{\alpha\beta\epsilon})+\frac{1}{2}R_{\gamma[\delta|\beta\epsilon}h_{|\alpha]\beta\epsilon}\nn\\
&-\frac{1}{8}h_{\delta\alpha\epsilon}F^{(-)j}_{\gamma\beta}F^{(-)j}_{\epsilon\beta}-\frac{1}{8}h_{\delta\alpha\epsilon}h_{\gamma\beta\omega}h_{\epsilon\beta\omega}-\frac{1}{16}h_{\alpha\beta\epsilon}F^{(-)\,j}_{\beta\epsilon}F^{(-)\,j}_{\gamma\delta}\nn\\
&+\frac{1}{8}F^{(-)\,j}_{\alpha\beta}\nabla^\beta F^{(-)\,j}_{\gamma\delta}-\frac{1}{8}h_{\delta\beta\epsilon}F^{(-)\,j}_{\beta\epsilon}F^{(-)\,j}_{\alpha\gamma}-\frac{1}{4}F^{(-)\,j}_{\beta\delta}\nabla^\beta F^{(-)\,j}_{\alpha\gamma}\nn\\
&+\frac{1}{4}h_{\beta\epsilon\delta}\nabla^\beta h_{\alpha\gamma\epsilon}-\nabla_{[\gamma}\mathcal E^{(0)}_{g,\delta]\alpha}-\frac{1}{2}\nabla_\gamma\mathcal E^{(0)}_{H,\delta\alpha}-\partial_\gamma\varphi\mathcal E^{(0)}_{H,\delta\alpha}\nn\\
&-\frac{1}{2}h_{\delta\alpha\epsilon}\mathcal E^{(0)}_{H,\gamma\epsilon}+\frac{1}{4}h_{\alpha\gamma\epsilon}\mathcal E^{(0)}_{H,\epsilon\delta}\Bigg]+R_{\alpha\beta}^{\ \ \ \gamma\delta}(\omega_+)\nabla^\beta F^{(-)\,i}_{\gamma\delta}.
\end{align}
We then substitute this back into $\mathcal E^{(0)}_{H,\alpha a}$, and take the difference $\qty(\delta \mathcal E^{(0)}_{H,\alpha a}+\mathcal E^{(1)}_{H,\alpha a})-2\qty(\delta \mathcal E^{(0)}_{g,\alpha a}+\mathcal E^{(1)}_{g,\alpha a})$. We then expand out the torsionful Riemann tensors as
\begin{equation}
    R_{\alpha\beta}^{\ \ \ \gamma\delta}(\omega_+)=R_{\alpha\beta}^{\ \ \ \gamma\delta}+\nabla_{[\alpha}h_{\beta]}^{\ \ \gamma\delta}+\frac{1}{2}h_{[\alpha}^{\ \ \gamma\epsilon}h_{\beta]}^{\ \ \epsilon\delta},
\end{equation}
and use the Riemann algebraic Bianchi identity $R_{[\alpha\beta\gamma]\delta}=0$, as well as the $h$ Bianchi identity \eqref{eq:hbian} and the $F$ Bianchi identity $\dd F^{(-)\,i}=0$, to get that
\begin{align}
    \qty(\delta \mathcal E^{(0)}_{H,\alpha a}+\mathcal E^{(1)}_{H,\alpha a})-2\qty(\delta \mathcal E^{(0)}_{g,\alpha a}+\mathcal E^{(1)}_{g,\alpha a})=&-\frac{1}{4}\delta_i^aF^{(-)\,i}_{\gamma\delta}\Bigg[-\nabla_{[\gamma}\mathcal E^{(0)}_{g,\delta]\alpha}-\frac{1}{2}\nabla_\gamma\mathcal E^{(0)}_{H,\delta\alpha}-\partial_\gamma\varphi\mathcal E^{(0)}_{H,\delta\alpha}\nn\\
    &-\frac{1}{2}h_{\delta\alpha\epsilon}\mathcal E^{(0)}_{H,\gamma\epsilon}+\frac{1}{4}h_{\alpha\gamma\epsilon}\mathcal E^{(0)}_{H,\epsilon\delta}\Bigg]+\delta e^a_i \delta^i_a\mathcal E^{(0)}_{H,\alpha a}-2\delta e^i_a\delta^a_i\mathcal E^{(0)}_{g,\alpha a},
\end{align}
which vanishes after the application of the two-derivative equations of motion. This verifies \eqref{eq:max2}. Hence, the truncation of the vector multiplet is consistent with the bosonic equations of motion.

\subsubsection{The surviving equations of motion}

The above shows that it is consistent to truncate away the bosonic equations of motion related to the reduced vector multiplet fields, namely $\mathcal E_{g,ij}$, $\mathcal E_{H,ij}$ and $\mathcal E_{H,\alpha i}-2\mathcal E_{g,\alpha i}$, corresponding to the equations of motion for $g_{ij}$, $b_{ij}$ and $F_{\mu\nu}^{(+)\,i}$, respectively.  The remaining untruncated equations of motion are those of the reduced supergravity multiplet fields.  These are the Einstein equation, $\mathcal E_{g,\alpha\beta}$, dilaton equation, $\mathcal E_\phi$, $h$-field equation, $\mathcal E_{H,\alpha\beta}$ and graviphoton equation, $-\mathcal E_{H,\alpha i}-2\mathcal E_{g,\alpha i}$.

Combined with the two-derivative equations of motion in (\ref{eq:2dtrunc}), the reduced Einstein equation is
\begin{align}
    \mathcal E_{g,\alpha\beta}&=R(\omega)_{\alpha\beta}-\fft14F_{\alpha\gamma}^{(-)\,i}F_{\beta\gamma}^{(-)\,i}-\fft14\tilde h_{\alpha\gamma\delta}\tilde h_{\beta\gamma\delta}+2\nabla_\alpha\nabla_\beta\varphi\nn\\
    &\quad+\frac{\alpha'}{4}\biggl(R_{\alpha\gamma\delta\epsilon}(\omega_+)R_\beta{}^{\gamma\delta\epsilon}(\omega_+)-R_{\alpha}{}^{\gamma\delta\epsilon}(\omega_+)F^{(-)\,i}_{\beta\gamma}F^{(-)\,i}_{\delta\epsilon}\nn\\
    &\kern4em+\frac{1}{8}F^{(-)\,i}_{\alpha\gamma}F_{\beta}{}^{\gamma(-)\,j}F^{(-)\,i}_{\delta\epsilon}F^{\delta\epsilon(-)\,j}+\frac{1}{4}\nabla^{(+)}_\alpha F^{(-)\,i}_{\gamma\delta}\nabla^{(+)}_
    \beta F^{\gamma\delta(-)\,i}\biggr),
\label{eq:redeins}
\end{align}
where symmetrization on $(\alpha\beta)$ is implicitly assumed.  Here we have introduced
\begin{equation}
    \tilde h=h-\fft{\alpha'}4\omega_{3L}(\omega_+)=\dd b+\fft14A^{(-)\,i}\wedge F^{(-)\,i}-\fft{\alpha'}4\omega_{3L}(\omega_+),
\label{eq:tildeH}
\end{equation}
such that
\begin{equation}
    \dd\tilde h=\fft14F^{(-)\,i}\wedge F^{(-)\,i}-\fft{\alpha'}4\Tr R(\omega_+)\wedge R(\omega_+).
\end{equation}
This reduced $\tilde h$ has the simple equation of motion
\begin{equation}
    \mathcal E_{H,\alpha\beta}=e^{2\varphi}\nabla^\gamma\qty(e^{-2\varphi}\tilde h_{\alpha\beta\gamma}).
\label{eq:redH}
\end{equation}

The remaining equations of motion are the dilaton equation
\begin{align}
    \mathcal E_\phi&=R(\omega)-\fft18F_{\alpha\beta}^{(-)\,i}F_{\alpha\beta}^{(-)\,i}-\fft1{12}\tilde h_{\alpha\beta\gamma}^2+4\Box\varphi-4(\partial_\alpha\varphi)^2,\nn\\
    &\quad+\fft{\alpha'}8\biggl((R_{\alpha\beta\gamma\delta}(\omega_+))^2-R_{\alpha\beta\gamma\delta}(\omega_+)F_{\alpha\beta}^{(-)\,i}F_{\gamma\delta}^{(-)\,i}+\fft12\qty(\nabla_\alpha^{(+)}F_{\beta\gamma}^{(-)\,i})^2\nn\\
    &\kern3em+\fft18F_{\alpha\beta}^{(-)\,i}F_{\beta\gamma}^{(-)\,i}F_{\gamma\delta}^{(-)\,j}F_{\delta\alpha}^{(-)\,j}-\fft18F_{\alpha\beta}^{(-)\,i}F_{\beta\gamma}^{(-)\,j}F_{\gamma\delta}^{(-)\,i}F_{\delta\alpha}^{(-)\,j}\nn\\
    &\kern3em+\fft18F_{\alpha\beta}^{(-)\,i}F_{\alpha\beta}^{(-)\,j}F_{\gamma\delta}^{(-)\,i}F_{\gamma\delta}^{(-)\,j}
    \biggr),
\label{eq:reddil}
\end{align}
and the graviphoton equation
\begin{align}
    \mathcal E_{A,\alpha i}&=e^{2\varphi}\nabla^\gamma\qty(e^{-2\varphi}F_{\gamma\alpha}^{(-)\,i})-\fft12\tilde h_{\alpha\beta\gamma}F_{\beta\gamma}^{(-)\,i}\nn\\
    &\quad+\fft{\alpha'}4\biggl(-\tilde h_{\alpha\beta\gamma}R_{\beta\gamma\delta\epsilon}(\omega_+)F_{\delta\epsilon}^{(-)\,i}+\fft14\tilde h_{\alpha\beta\gamma}F_{\beta\gamma}^{(-)\,j}F_{\delta\epsilon}^{(-)\,i}F_{\delta\epsilon}^{(-)\,j}-2R_{\alpha\beta\gamma\delta}(\omega_+)\nabla_\beta^{(+)}F_{\gamma\delta}^{(-)\,i}\nn\\
    &\kern4em+\fft12F_{\alpha\gamma}^{(-)\,j}F_{\delta\epsilon}^{(-)\,j}\nabla_\gamma^{(+)}F_{\delta\epsilon}^{(-)\,i}-\fft12F_{\alpha\gamma}^{(-)\,j}F_{\delta\epsilon}^{(-)\,i}\nabla_\gamma^{(+)}F_{\delta\epsilon}^{(-)\,j}\nn\\
    &\kern4em+F_{\beta\gamma}^{(-)\,i}F_{\gamma\delta}^{(-)\,j}\nabla_\alpha^{(+)}F_{\beta\delta}^{(-)\,j}\biggr).
\label{eq:redA}
\end{align}
Note that, at the order we are working at, we cannot distinguish between $h$ and $\tilde h$ in the $\mathcal O(\alpha')$ terms.  Nevertheless, the use of $\tilde h$ is expected to be natural when extending to additional higher orders in $\alpha'$.

\subsection{The reduced Lagrangian}

Since a torus reduction is known to be consistent, the reduced equations of motion will necessarily be consistent with the Kaluza-Klein reduced Lagrangian.  With the further consistent truncation of the vector multiplets, the supergravity multiplet equations of motion, (\ref{eq:redeins}), (\ref{eq:redH}), (\ref{eq:reddil}) and (\ref{eq:redA}), may be derived from the effective Lagrangian given by substituting the truncation into our Lagrangian \eqref{eq:Lhet}, which gives
\begin{align}
    e^{-1}\mathcal L=&e^{-2\varphi}\biggl[R+4\qty(\partial\varphi)^2-\fft1{12}\tilde h_{\mu\nu\rho}^2-\fft18\qty(F_{\mu\nu}^{(-)\,i})^2\nn\\
    &\qquad+\frac{\alpha'}{8}\biggl(\qty(R_{\mu\nu\rho\sigma}(\omega_+))^2-R^{\mu\nu\rho\sigma}(\omega_+)F_{\mu\nu}^{(-)\,i}F_{\rho\sigma}^{(-)\,i} +\frac{1}{2}\qty(\nabla_\rho^{(+)}F_{\mu\nu}^{(-)\,i})^2\nn\\
    &\kern5em+\frac{1}{8}F_\mu{}^{\nu\,(-)\,i}F_\nu{}^{\rho(-)\,i}F_\rho{}^{\sigma\,(-)\,j}F_\sigma{}^{\mu\,(-)\,j}-\frac{1}{8}F_\mu{}^{\nu\,(-)\,i}F_\nu{}^{\rho(-)\,j}F_\rho{}^{\sigma\,(-)\,i}F_\sigma{}^{\mu\,(-)\,j}\nn\\
    &\kern5em+\frac{1}{8}F_{\mu\nu}^{(-)\,i}F^{\mu\nu\,(-)\,j}F^{(-)\,i}_{\rho\sigma}F^{\rho\sigma\,(-)\,j}\biggr)\biggr].
\end{align}
While higher-derivative Lagrangians can be transformed by field redefinitions, it is important to note that such field redefinitions will also transform the equations of motion.  This form of the Lagrangian is what matches the equations of motion given above.

By making use of our freedom to perform field redefinitions, we may transform this Lagrangian into a more standard form.  In particular, we can rewrite the $(\nabla F)^2$ term using the identity
\begin{align}
    \qty(\nabla_\rho^{(+)}F_{\mu\nu}^{(-)\,i})^2&=R^{\mu\nu\rho\sigma}(\omega_+)F_{\mu\nu}^{(-)\,i}F_{\rho\sigma}^{(-)\,i}-\fft14F_{\mu\nu}^{(-)\,i}F^{\mu\nu\,(-)\,j}F^{(-)\,i}_{\rho\sigma}F^{\rho\sigma\,(-)\,j}\nn\\
    &\quad-\fft12F_\mu{}^{\nu\,(-)\,i}F_\nu{}^{\rho(-)\,i}F_\rho{}^{\sigma\,(-)\,j}F_\sigma{}^{\mu\,(-)\,j}+\fft12F_\mu{}^{\nu\,(-)\,i}F_\nu{}^{\rho(-)\,j}F_\rho{}^{\sigma\,(-)\,i}F_\sigma{}^{\mu\,(-)\,j}\nn\\
    &\quad+2e^{2\varphi}\nabla_\rho(e^{-2\varphi}F_{\mu\nu}^{(-)\,i}\nabla^\mu F^{\rho\nu\,(-)\,i})-2F^{\mu\nu\,(-)\,i}\nabla_\mu\mathcal E_\nu^{(0)\,i}-2F_{\mu\rho}^{(-)\,i}F_\nu{}^{\rho\,(-)\,i}\mathcal E_{g,\mu\nu}^{(0)}.
\end{align}
Here, the last line includes a total derivative along with terms proportional to the leading order equations of motion.  The equation of motion terms can be removed by a suitable field redefinition, in which case we end up with the concise expression
\begin{align}
    e^{-1}\mathcal L=&e^{-2\varphi}\biggl[R+4\qty(\partial\varphi)^2-\fft1{12}\tilde h_{\mu\nu\rho}^2-\fft18\qty(F_{\mu\nu}^{(-)\,i})^2\nn\\
    &\qquad+\frac{\alpha'}{8}\biggl(\qty(R_{\mu\nu\rho\sigma}(\omega_+))^2-\fft12R^{\mu\nu\rho\sigma}(\omega_+)F_{\mu\nu}^{(-)\,i}F_{\rho\sigma}^{(-)\,i} \nn\\
    &\kern5em-\frac{1}{8}F_\mu{}^{\nu\,(-)\,i}F_\nu{}^{\rho(-)\,i}F_\rho{}^{\sigma\,(-)\,j}F_\sigma{}^{\mu\,(-)\,j}+\frac{1}{8}F_\mu{}^{\nu\,(-)\,i}F_\nu{}^{\rho(-)\,j}F_\rho{}^{\sigma\,(-)\,i}F_\sigma{}^{\mu\,(-)\,j}\biggr)\biggr].
\label{eq:Ltruncred}
\end{align}

This reduced Lagrangian can be compared with that of Ref.~\cite{Eloy:2020dko}.  Note, however, that Ref.~\cite{Eloy:2020dko} does not use a torsionful connection.  In order to make the comparison, we may use the identities
\begin{align}
    R_{\mu\nu\rho\sigma}(\omega_+)^2&=R_{\mu\nu\rho\sigma}^2+\fft12R^{\mu\nu\rho\sigma}h_{\mu\nu\lambda}h_{\rho\sigma}{}^\lambda-\fft18(h^2_{\mu\nu})^2-\fft18h^4+\fft3{32}F_{\mu\nu}^{(-)\,i}F^{\mu\nu\,(-)\,j}F^{(-)\,i}_{\rho\sigma}F^{\rho\sigma\,(-)\,j}\nn\\
    &\quad-\fft3{16}F_\mu{}^{\nu\,(-)\,i}F_\nu{}^{\rho(-)\,j}F_\rho{}^{\sigma\,(-)\,i}F_\sigma{}^{\mu\,(-)\,j}-\fft14h^{2\,\mu\nu}F_{\mu\rho}^{(-)\,i}F_\nu{}^{\rho\,(-)\,i}\nn\\
    &\quad+\fft18h^{\mu\nu\lambda}h^{\rho\sigma}{}_\lambda F_{\mu\nu}^{(-)\,i}F_{\rho\sigma}^{(-)\,i}-\fft14h^{\mu\nu\lambda}h^{\rho\sigma}{}_\lambda F_{\mu\rho}^{(-)\,i}F_{\nu\sigma}^{(-)\,i}\nn\\
    &\quad+e^{2\varphi}\nabla_\mu(e^{-2\varphi}h_{\nu\rho\sigma}\nabla^\nu h^{\mu\rho\sigma})-h^{\nu\rho\sigma}\nabla_\nu\mathcal E_{H,\rho\sigma}^{(0)}-h^{2\,\mu\nu}\mathcal E_{g,\mu\nu}^{(0)},
\end{align}
and
\begin{align}
    R^{\mu\nu\rho\sigma}(\omega_+)F_{\mu\nu}^{(-)\,i}F_{\rho\sigma}^{(-)\,i}&=R^{\mu\nu\rho\sigma}F_{\mu\nu}^{(-)\,i}F_{\rho\sigma}^{(-)\,i}+\fft18F_{\mu\nu}^{(-)\,i}F^{\mu\nu\,(-)\,j}F^{(-)\,i}_{\rho\sigma}F^{\rho\sigma\,(-)\,j}\nn\\
    &\quad-\fft14F_\mu{}^{\nu\,(-)\,i}F_\nu{}^{\rho(-)\,j}F_\rho{}^{\sigma\,(-)\,i}F_\sigma{}^{\mu\,(-)\,j}-\fft12h^{\mu\nu\lambda}h^{\rho\sigma}{}_\lambda F_{\mu\rho}^{(-)\,i}F_{\nu\sigma}^{(-)\,i},
\end{align}
where we have defined $h^4=h^{\mu\nu\rho}h_{\mu\sigma\lambda}h_{\nu\sigma\epsilon}h_{\rho\lambda\epsilon}$. Note that $\tilde h$ defined in (\ref{eq:tildeH}) also contributes to the four derivative action through $\tilde h_{\mu\nu\rho}^2$.  At $\mathcal O(\alpha')$ we only need to worry about the cross-term
\begin{align}
    h^{\mu\nu\rho}\omega_{3L\,\mu\nu\rho}(\omega_+)&=h^{\mu\nu\rho}\omega_{3L\,\mu\nu\rho}-3R^{\mu\nu\rho\sigma}h_{\mu\nu\lambda}h_{\rho\sigma}{}^\lambda-\fft3{16}h^{\mu\nu\lambda}h^{\rho\sigma}{}_\lambda F_{\mu\nu}^{(-)\,i}F_{\rho\sigma}^{(-)\,i}\nn\\
    &\quad+\fft38h^{\mu\nu\lambda}h^{\rho\sigma}{}_\lambda F_{\mu\rho}^{(-)\,i}F_{\nu\sigma}^{(-)\,i}+\fft12h^4.
\end{align}
Making the above substitutions in (\ref{eq:Ltruncred}) then gives us
\begin{align}
    e^{-1}\mathcal L=&e^{-2\varphi}\biggl[R+4\qty(\partial\varphi)^2-\fft1{12} h_{\mu\nu\rho}^2-\fft18\qty(F_{\mu\nu}^{(-)\,i})^2\nn\\
    &+\frac{\alpha'}{8}\biggl(\qty(R_{\mu\nu\rho\sigma})^2-\frac{1}{2}R_{\mu\nu\rho\sigma}h^{\mu\nu\lambda}h^{\rho\sigma}{}_{\lambda}-\frac{1}{2}R^{\mu\nu\rho\sigma}F^{(-)\,i}_{\mu\nu}F^{(-)\,i}_{\rho\sigma}-\frac{1}{8}(h^2_{\mu\nu})^2+\frac{1}{24}h^4\nn\\
    &-\frac{1}{4}h^2_{\mu\nu} F^{(-)\,i}_{\mu\rho}F^{(-)\,i}_{\nu\rho}+\frac{1}{16}h^{\mu\nu\lambda}h^{\rho\sigma}{}_{\lambda}F^{(-)\,i}_{\mu\nu}F^{(-)\,i}_{\rho\sigma}+\frac{1}{8}h^{\mu\nu\lambda}h^{\rho\sigma}{}_\lambda F^{(-)\,i}_{\mu\rho}F^{(-)\,i}_{\nu\sigma}\nn\\
    &+\frac{1}{32}F^{(-)\,i}_{\mu\nu}F^{(-)\,j}_{\mu\nu}F^{(-)\,i}_{\rho\sigma}F^{(-)\,j}_{\rho\sigma}-\frac{1}{8}F^{(-)\,i}_{\mu\nu}F^{(-)\,i}_{\nu\sigma}F^{(-)\,j}_{\sigma\rho}F^{(-)\,j}_{\rho\mu}+\frac{1}{16}F^{(-)\,i}_{\mu\nu}F^{(-)\,j}_{\nu\sigma}F^{(-)\,i}_{\sigma\rho}F^{(-)\,j}_{\rho\mu}\nn\\
    &+\fft13h^{\mu\nu\rho}\omega_{3L\,\mu\nu\rho}\biggr)\biggr].
\end{align}
In particular, this agrees with the result of \cite{Eloy:2020dko} after appropriate truncation\footnote{It is important to note that our conventions differ from those of \cite{Eloy:2020dko}. In particular, one must send $B\to-B$ to compare, so our truncation $F^{(+)}=0$ is equivalent to $F^{(-)}=0$ in the notation of \cite{Eloy:2020dko}.}.

\section{Truncating the fermionic sector}\label{sec:fermionic}

Hitherto, we have only looked at the bosonic sector of the theory; it is a non-trivial test to additionally check that the truncation extends to the fermion sector.  We start with the $\mathcal O(\alpha')$ truncation
\begin{equation}
    g_{ij}=\delta_{ij}+\alpha'\delta g_{ij},\qquad b_{ij}=0,\qquad F_{\mu\nu}^i=\fft12F_{\mu\nu}^{(-)\,i},\qquad\tilde G_{\mu\nu\,i}=-\fft12F_{\mu\nu}^{(-)\,i},
\label{eq:truncanz}
\end{equation}
and consider the leading order supersymmetry variations, (\ref{eq:deltalo}).  We find, at the lowest order
\begin{align}
    \delta_\epsilon\psi_\mu^{(0)}&=\left(\nabla_\mu(\omega_-)+\fft14F_{\mu\nu}^{(-)\,i}\gamma^\nu\Gamma^{\underline i}\right)\epsilon,\nn\\
    \delta_\epsilon\psi_i^{(0)}&=0,\nn\\
    \delta_\epsilon\tilde\lambda^{(0)}&=\left(\gamma^\mu\partial_\mu\varphi-\fft1{12}h_{\mu\nu\lambda}\gamma^{\mu\nu\lambda}+\fft18F_{\mu\nu}^{(-)\,i}\gamma^{\mu\nu}\Gamma^{\underline i}\right)\epsilon,
\label{eq:dello}
\end{align}
where the internal Dirac matrices $\Gamma^{\underline i}$ have flat-space indices.  We see that the gaugini are consistently truncated out at this order.  However, with the $\delta g_{ij}$ shift in (\ref{eq:truncanz}), the lowest order transformations, (\ref{eq:deltalo}), also give rise to the $\mathcal O(\alpha')$ terms
\begin{align}
    \delta(\delta_\epsilon\psi_\mu^{(0)})&=0,\nn\\
    \delta(\delta_\epsilon\psi_i^{(0)})&=\left(-\fft1{16}\delta g_{ij}F_{\mu\nu}^{(-)\,j}\gamma^{\mu\nu}-\fft14\partial_\mu\delta g_{ij}\gamma^\mu\Gamma^{\underline{j}}\right)\epsilon,\nn\\
    \delta(\delta_\epsilon\tilde\lambda^{(0)})&=0.
\label{eq:3dfromlo}
\end{align}

\subsection{The variations at \texorpdfstring{$\mathcal O(\alpha')$}{O(alpha prime)}}

The shift in the lowest order internal gravitino variation, $\delta(\delta_\epsilon\psi_i^{(0)})$,  will combine with the higher order term, $\delta_\epsilon\psi_i^{(1)}$, to yield the complete gaugino variation.  As we are aiming to truncate away the vector multiplets, this combined variation ought to vanish.

Reduction of the first order internal gravitino variation, $\delta\psi_i^{(1)}$, in (\ref{eq:deltas}) yields
\begin{align}
    \delta_\epsilon\psi_i^{(1)}&=\fft1{32}\Bigl(R_{\alpha\beta}{}^{\gamma\delta}(\omega_+)F_{\gamma\delta}^{(-)\,i}\gamma^{\alpha\beta}-\fft18F_{\alpha\beta}^{(-)\,j}F_{\gamma\delta}^{(-)\,j}F_{\gamma\delta}^{(-)\,i}\gamma^{\alpha\beta}+F_{\gamma\delta}^{(-)\,[j}\nabla_\alpha^{(+)}F_{\gamma\delta}^{(-)\,i]}\gamma^\alpha\Gamma^{\underline j}\nn\\
    &\qquad\qquad-\fft12F_{\alpha\beta}^{(-)\,i}F_{\beta\gamma}^{(-)\,j}F_{\gamma\alpha}^{(-)\,k}\Gamma^{\underline{jk}}\Bigr)\epsilon.
\end{align}
Combining this with (\ref{eq:3dfromlo}) gives
\begin{align}
    \delta(\delta_\epsilon\psi_i^{(0)})+\delta_\epsilon\psi_i^{(1)}&=\fft1{32}F_{\alpha\beta}^{(-)\,i}\bigg(R_{\gamma\delta}{}^{\alpha\beta}(\omega_+)\gamma^{\gamma\delta}-\fft14F_{\alpha\beta}^{(-)\,j}F_{\gamma\delta}^{(-)\,j}\gamma^{\gamma\delta}-\nabla_\gamma^{(+)} F_{\alpha\beta}^{(-)\,j}\gamma^\gamma\Gamma^{\underline j}\nn\\
    &\qquad\qquad\qquad-\fft12F_{\beta\gamma}^{(-)\,j}F_{\gamma\alpha}^{(-)\,k}\Gamma^{\underline{jk}}\bigg)\epsilon.
\label{eq:delgau}
\end{align}
As this is non-vanishing, the gaugino must be shifted if we are to truncate it away.

The Riemann tensor term in the variation suggests that we use the commutator of two covariant derivatives.  This can arise from the variation of a covariant derivative of the gravitino.  Since we work only to $\mathcal O(\alpha')$, we can take something like $\nabla_{[\mu}\psi_{\nu]}$ whose variation will gives $\nabla_{[\mu}\nabla_{\nu]}\epsilon$.  To be more precise, consider the variation $\delta\psi_\mu^{(0)}$ in (\ref{eq:dello}), which we write as $\delta\psi_\mu^{(0)}=\mathcal D_\mu\epsilon$ where
\begin{equation}
    \mathcal D_\mu\equiv\nabla_\mu(\omega_-)+\fft14F_{\mu\nu}^{(-)\,i}\gamma^\nu\Gamma^{\underline i}=\nabla_\mu-\fft18h_{\mu\nu\lambda}\gamma^{\nu\lambda}+\fft14F_{\mu\nu}^{(-)\,i}\gamma^\nu\Gamma^{\underline i}.
\end{equation}
The commutator we want is then
\begin{align}
    [\mathcal D_\mu,\mathcal D_\nu]=&\fft14\bigg(R_{\mu\nu}{}^{\alpha\beta}(\omega_-)\gamma^{\alpha\beta}+\left(2\nabla_{[\mu}F_{\nu]\alpha}^{(-)\,i}-h_{[\mu}{}^{\alpha\beta}F_{\nu]\beta}^{(-)\,i}\right)\gamma^\alpha\Gamma^{\underline i}-\fft12F_{\mu\alpha}^{(-)\,i}F_{\nu\beta}^{(-)\,i}\gamma^{\alpha\beta}\nn\\
    &\qquad-\fft12F_{\mu\alpha}^{(-)\,i}F_{\nu\alpha}^{(-)\,j}\Gamma^{\underline{ij}}\bigg).
\end{align}
Comparing this with (\ref{eq:delgau}) indicates that we need to convert between $R(\omega_+)$ and $R(\omega_-)$.  Using the $h$ Bianchi identity, $\dd h=\fft14F^{(-)\,i}\wedge F^{(-)\,i}$, we have
\begin{equation}
    R_{\alpha\beta\gamma\delta}(\omega_-)=R_{\gamma\delta\alpha\beta}(\omega_+)-\fft34F_{[\alpha\beta}^{(-)\,i}F_{\gamma\delta]}^{(-)\,i}.
\end{equation}
A bit of manipulation, including use of the $F$ Bianchi $\dd F^{(-)\,i}=0$, then shows that $\delta_\epsilon\tilde\psi_i=0$ where we have defined the shifted gaugino
\begin{equation}
    \tilde\psi_i=\psi_i-\frac{\alpha'}{4}F_{\mu\nu}^{(-)\,i}\mathcal D_\mu\psi_\nu.
\end{equation}
This then demonstrates that we can consistently truncate $\tilde\psi_i$ out of the fermion sector while preserving the supersymmetry of the solution.

For completeness, we also note that the gravitino and dilatino variations have four-derivative corrections
\begin{align}
    \delta_\epsilon\psi_\mu^{(1)}=&\frac{1}{32}\Bigg[\omega_{3L,\mu\nu\rho}(\omega_+)\gamma^{\nu\rho}+2\qty(R_{\mu\nu}{}^{\alpha\beta}(\omega_+)-\frac{1}{8}F^{(-)\,j}_{\mu\nu}F^{(-)\,j}_{\alpha\beta})F_{\alpha\beta}^{(-)\,i}\gamma^\nu\Gamma^{\underline i}\nn\\
    &\qquad+\frac{1}{2}F^{(-)\,i}_{\alpha\beta}\nabla^{(+)}_\mu F^{(-)\,j}_{\alpha\beta}\Gamma^{\underline{ij}}\Bigg]\epsilon,
\end{align}
and
\begin{align}
    \delta_\epsilon\tilde\lambda^{(1)}&=\frac{1}{48}\biggl[\omega_{3L,\mu\nu\rho}(\omega_+)\gamma^{\mu\nu\rho}+\frac{3}{2}\left(R_{\mu\nu}{}^{\alpha\beta}(\omega_+)-\fft18F_{\mu\nu}^{(-)\,j}F_{\alpha\beta}^{(-)\,j}\right)F_{\alpha\beta}^{(-)\,i}\gamma^{\mu\nu}\Gamma^{\underline{i}}\nn\\
    &\kern4em+\fft14F^{(-)\,i}_{\alpha\beta}F^{(-)\,j}_{\beta\gamma}F^{(-)\,k}_{\gamma\alpha}\Gamma^{\underline{ijk}}\biggr]\epsilon.
\end{align}
%
%

\section{Comparing with the four-derivative corrected BPS black string}\label{sec:example}

A natural test for our truncation is to consider $\mathcal N=1$, $D=9$ supergravity \cite{Gates:1984kr}. The field content is the 9D metric $g_{\mu\nu}$, the gravitino $\psi_\mu$, a vector $A_\mu$, a two-form $B_{\mu\nu}$, a dilatino $\lambda$, and a dilaton $\varphi$.  Black hole solutions to this minimal supergravity lift to ten-dimensional strings, so we may compare between these two cases.

\subsection{Two-derivative solution}

Balancing the nine-dimensional graviphoton charge with the mass gives rise to a supersymmetric black hole solution given by \cite{Lu:1995cs}
\begin{align}
    \dd s_9^2&=-\qty(1+\frac{k}{r^6})^{-2}\dd t^2+\dd r^2+r^2\dd\Omega_{7}^2,\nn\\
    A&=\frac{1}{1+\frac{k}{r^6}}\dd t,\nn\\
    e^{\varphi}&=\qty(1+\frac{k}{r^6})^{-1/2},
\end{align}
with all other fields vanishing. Here we have denoted the metric of the round $S^7$ by $\dd\Omega_7^2$. This two-derivative solution uplifts to a 10D black string solution via \eqref{eq:redMet}
\begin{align}
    \dd s_{10}^2&=-\qty(1+\frac{k}{r^6})^{-2}\dd t^2+\dd r^2+r^2\dd\Omega_{7}^2+\qty(\dd z+\frac{\dd t}{1+\frac{k}{r^6}})^2,\nn\\
    B&=-\frac{1}{1+\frac{k}{r^6}}\dd t\land \dd z,\nn\\
    e^{\phi}&=\qty(1+\frac{k}{r^6})^{-1/2}.
\end{align}
It is straightforward to check that this satisfies the 10D two-derivative equations of motion \eqref{eq:eom2dred}. Moreover, this is a BPS solution, in the sense that $\delta\psi_M^{(0)}=0$ and $\delta\lambda^{(0)}=0$ with Killing spinor
\begin{equation}
    \epsilon^{(0)}=\qty(1+\frac{k}{r^6})^{-1/2}\qty(1-\gamma^{0}\Gamma^{\underline z})\epsilon_0(\Omega_7),
\end{equation}
where $\epsilon_0$ is a covariantly constant spinor\footnote{Concretely, if we write the seven angles of $S^7$ as $\theta_n$, $n=1,...,7$, such that the angles are defined recursively as $\dd \Omega_{n}^2=\dd\theta_{8-n}^2+\sin^2\theta_{8-n}\dd \Omega_{{n-1}}^2$, then $\epsilon_0(\Omega_7)$ is given by $\epsilon_0=\prod_{n=1}^7\exp[\frac{\theta_n}{2}\gamma^{n+1,n+2}]\eta_0$, where $\eta_0$ is a constant spinor, \textit{i.e.}, with no coordinate dependence. This can be seen via an identical analysis to that in \cite{Lu:1998nu}.} on $S^7$. Thus, we see the solution is $\frac{1}{2}$-BPS. 

\subsection{Four-derivative correction}

We now solve the ten-dimensional equations of motion including $\mathcal O(\alpha')$ corrections.  We do so by making an ansatz for the four-derivative corrected black string as
\begin{align}
    \dd s_{10}^2=&-\qty(1+\frac{k}{r^6})^{-2}\qty(1+\alpha' f(r))\dd t^2+\qty(1+\alpha' g(r))\dd r^2+r^2\dd\Omega_{7}^2\nn\\
    &+\qty(1+\alpha' h(r))\qty(\dd z+\qty(1+\alpha' j(r))\frac{\dd t}{1+\frac{k}{r^6}})+\mathcal{O}(\alpha'^2),\nn\\
    B=&-\frac{1+\alpha' \kay(r)}{1+\frac{k}{r^6}}\dd t\land \dd z+\mathcal{O}(\alpha'^2),\nn\\
    e^{\phi}=&\qty(1+\alpha' \ell(r))\qty(1+\frac{k}{r^6})^{-1/2}+\mathcal{O}(\alpha'^2),
\end{align}
where we have explicitly assumed that the $SO(8)$ symmetry is preserved and that the solution continues to have $\partial_t$ and $\partial_z$ as Killing vectors. Demanding that our fermion variations vanish implies the particularly useful conditions
\begin{align}
    0&=g(r),\nn\\
    0&=\dv{}{r}\qty(\frac{f(r)-h(r)-2j(r)}{1+\frac{k}{r^6}}),\nn\\
    0&=\dv{}{r}\qty(\frac{h(r)+j(r)-\kay(r)}{1+\frac{k}{r^6}}),\nn\\
    0&=6kf(r)-\qty(1+\frac{k}{r^6})^2\dv{}{r}\qty(\frac{j(r)+\kay(r)}{1+\frac{k}{r^6}})+4r(k+r^6)m'(r),\nn\\
    0&=3k\qty(2\kay(r)-f(r)-h(r))+r(k+r^6)\qty(\kay'(r)-2\ell'(r)),\label{eq:BPScond1}
\end{align}
where we have written our BPS spinor as $\epsilon=(1+\alpha'm(r))\epsilon^{(0)}$. Note that since we are demanding $\delta\psi_\mu=0$ and $\delta\lambda=0$ for a BPS solution, it is fine to require that $\delta\psi_z=0$ without any field redefinitions. In particular, \eqref{eq:BPScond1} allows us to write
\begin{align}
    f(r)-h(r)-2j(r)=&c_1\qty(1+\frac{k}{r^6}),\nn\\
    h(r)+j(r)-\kay(r)=&-c_2\qty(1+\frac{k}{r^6}),\nn\\
    \kay'(r)-2\ell'(r)=&\frac{3k(c_1-2c_2)}{r(k+r^6)}\qty(1+\frac{k}{r^6}),\label{eq:BPScond2}
\end{align}
where $c_1$ and $c_2$ are undetermined constants of integration. Substituting these conditions \eqref{eq:BPScond2} into our equations of motion and solving gives the solution
\begin{align}
    f(r)=&\frac{18 k^2}{r^2(k+r^6)^2}+c_4+2c_7+c_1\qty(2-\frac{1}{1+\frac{k}{r^6}})-\frac{c_3+2c_6}{6(k+r^6)},\nn\\
    g(r)=&0,\nn\\
    h(r)=&-\frac{18 k^2}{r^2(k+r^6)^2}-\frac{c_3}{6(k+r^6)}+c_4,\nn\\
    j(r)=&\frac{18 k^2}{r^2(k+r^6)^2}-\frac{c_1k^2}{2r^6(k+r^6)}-\frac{c_6}{6(k+r^6)},\nn\\
    \kay(r)=&\frac{(c_1-2c_2)k}{2r^6}+\frac{3c_1k-c_3-c_6}{6(k+r^6)}+c_2+c_4+c_7,\nn\\
    \ell(r)=&\frac{3c_1k-c_3-c_6}{12(k+r^6)}+c_5+\frac{c_7}{2},
\end{align}
where the $c_i$ are (as yet undetermined) constants of integration, along with the four-derivative corrected Killing spinor
\begin{equation}
    \epsilon=\qty[1+\frac{\alpha'}{2}\qty(\frac{9k^2}{r^2(k+r^6)^2}-\frac{c_3+2c_6-6c_1 k}{12(k+r^6)})]\epsilon^{(0)}.
\end{equation}
Now, we can further reduce the number of free constants. Our mass $M$, electric charge $Q^{(A)}$, $B$ charge $Q^{(B)}$, and scalar charge $\Sigma$ are given by
\begin{align}
    M&=2k+\alpha'\qty(-\qty(3c_1+2c_4+4c_7)k+\frac{c_3+2c_6}{6}),\nn\\
    Q^{(A)}&=-k-\alpha'\qty(\frac{c_6}{6}+c_7k),\nn\\
    Q^{(B)}&=k+\alpha'\qty(\frac{c_3+c_6}{6}+(c_4+c_7)k),\nn\\
    \Sigma&=-\frac{k}{2}+\frac{\alpha'}{12}\qty(3k\qty(c_1-2c_5-c_7)-c_3-c_6),
\end{align}
where the scalar charge is defined as the coefficient of $r^{-6}$ in the large-$r$ expansion of the dilaton
\begin{equation}
    \phi=\phi_{\infty}+\frac{\Sigma}{r^6}+\mathcal{O}(r^{-12}).
\end{equation}
If we fix the metric at infinity, as well as the asymptotic value of the scalar $\phi_\infty$ and the charges $Q^{(A)}$ and $Q^{(B)}$, then this requires that all the $c_i$ vanish. Upon doing so, we are left with the four-derivative corrected metric
\begin{align}
    \dd s_{10}^2=&-\qty(1+\frac{k}{r^6})^{-2}\qty(1+\frac{18\alpha' k^2}{r^2(k+r^6)^2})\dd t^2+\dd r^2+r^2\dd\Omega_7^2\nn\\
    &+\qty(1-\frac{18\alpha' k^2}{r^2(k+r^6)^2})\qty(\dd z-\frac{1}{1+\frac{k}{r^6}}\qty(1+\frac{18\alpha' k^2}{r^2(k+r^6)^2})\dd t)^2+\mathcal{O}(\alpha'^2),
\end{align}
with $B$ and $\phi$ left uncorrected.

We now compare this ten-dimensional solution with the form of the consistent truncation, (\ref{eq:truncanz1}).  Making note of
\begin{equation}
    F^2=\frac{1}{3}H^2=-\frac{72k^2}{r^2(k+r^6)^2},
\end{equation}
we find
\begin{subequations}
   \begin{align}
    g_{zz}=& 1+\frac{\alpha'}{4}F^2,\label{eq:gzz}\\
    B =& \qty(-A+\frac{\alpha'}{2}\Omega^{\alpha\beta}F_{\alpha\beta})\land\dd z.\label{eq:B}
\end{align}
\end{subequations}
In particular, \eqref{eq:gzz} is exactly what we expect from the general expression \eqref{eq:truncanz1} when $n=1$. On the other hand, \eqref{eq:B} na\"ively seems to have an extra term compared with \eqref{eq:truncanz}, but the $\Tr\Omega F$ is precisely the term that we implicitly absorbed into $B$ to remove a total derivative from the Lorentz Chern-Simons form.

\section{Discussion}\label{sec:conclusion}

We have shown that it is consistent to truncate out the vector multiplets that arise in the toroidal reduction of heterotic supergravity in the presence of four-derivative corrections. In particular, this truncation does not ruin the supersymmetry of the reduced theory. We further verified our truncation by looking at the example of a four-derivative corrected black string solution. We view this work as a step towards more general non-trivial truncations of higher-derivative theories, such as what may arise from sphere reductions.

One may be tempted to interpret these results as a statement that a two-derivative truncation automatically implies the existence of a four-derivative one. However, this is not always the case. For example, one might consider further truncating the $T^5$ reduction to minimal $D=5$, $\mathcal{N}=2$ supergravity. For $n=5$, upon transforming to the Einstein frame $g=e^{4\varphi/3}\tilde g$ and dualizing $h=e^{-2\varphi}\star \mathcal G=e^{-2\varphi}\star\dd C$, the two-derivative Lagrangian reads\footnote{Here, we have chosen to use the notation $\star$ to refer to the Hodge star in the string frame and $\tilde\star$ the Hodge star in the Einstein frame.},
\begin{equation}
    \tilde\star\mathcal L^{(0)}=\tilde\star R-\frac{8}{3}\dd\varphi\land\tilde\star\dd\varphi-\frac{1}{2}e^{8\varphi/3}\mathcal G\land\tilde\star\mathcal G-\frac{1}{4}F^{(-)\,i}\land F^{(-)\,i}\land C-\frac{1}{4}e^{-4\varphi/3}F^{(-)\,i}\land\tilde\star F^{(-)\,i},
\end{equation}
where the Chern-Simons term arises from requiring that the $h$ Bianchi identity become the $\mathcal G$ equation of motion. This theory may be thought of, in bosonic $\mathcal{N}=2$ language, as a graviton multiplet $(g_{\mu\nu},C_{\mu})$ coupled to a gravitino multiplet $A^{(-)\,i\ne 1}_{\mu}$ and a vector multiplet $(A^{(-)\,1}_{\mu},\varphi)$. One may then check that it is consistent to truncate the additional multiplets
\begin{equation}
    F^{(-)\,1}=\pm 2\mathcal G,\qquad F^{(-)\,i\ne 1}=0,\qquad \varphi = 0,
\end{equation}
which, upon rescaling $\mathcal G=F/\sqrt{3}$, gives us pure minimal ungauged $\mathcal N=2$ supergravity
\begin{equation}
    \star\mathcal L^{(0)}=\star R-\frac{1}{2}F\land\star F-\frac{1}{3\sqrt{3}}F\land F\land A.
\end{equation}
Similarly, upon truncating, the two-derivative dilatino equation takes the form
\begin{equation}
    \delta_\epsilon\tilde\lambda^{(0)}=-\frac{i}{4\sqrt{3}}F_{\mu\nu}\gamma^{\mu\nu}\qty(1\pm i\Gamma^{\underline 1})\epsilon,
\end{equation}
where the presence of a projector is consistent with the fact that we are truncating $\mathcal N=4$ supersymmetry down to $\mathcal{N}=2$.

At the four-derivative level, we na\"ively expect the minimal $\mathcal N=2$ truncation to yield the Lagrangian of \cite{Bobev:2021qxx,Liu:2022sew,Bobev:2022bjm}. However, the dilaton equation of motion in the Einstein frame contains the term
\begin{equation}
    \mathcal E_\varphi^{(1)}\supset \frac{1}{3}e^{4\varphi/3}\qty(R_{\mu\nu\rho\sigma})^2.
\end{equation}
As is well-known, such a term cannot be removed by field redefinitions. This spells doom for our truncation. An alternative way of seeing the same issue is to look at the $\mathcal G$ equation of motion (or, equivalently, the $\tilde h$ Bianchi identity)
\begin{equation}
    \dd\qty(e^{-2\varphi}\star \mathcal G)\sim -\frac{\alpha'}{4}R(\omega_+)\land R(\omega_+)+...
\end{equation}
No such term appears in the $F^{(-)\,i}$ equation of motion, and likewise cannot be removed by a field redefinition. The issue in both cases is that the two-derivative equations of motion have no Riemann tensors and so field redefinitions cannot generate two Riemann tensors%
\footnote{Although, if we are clever, we can generate one, as happened for the Maxwell equation truncation earlier.}.
Hence, the truncation is likely inconsistent at the four-derivative level. We may also see this in the fermionic sector. If we now identify $\tilde h=e^{-2\varphi}\star\mathcal G$ and set $F^{(-)\,i\ne 0}=0$, then the four-derivative part of the dilatino variation becomes
\begin{equation}
    \delta_\epsilon\tilde\lambda^{(1)}=\frac{1}{32}\qty(R_{\mu\nu}{}^{\alpha\beta}(\omega_+)-\frac{1}{8}F^{(-)\,1}_{\mu\nu}F^{(-)\,1}_{\alpha\beta})F^{(-)\,1}_{\alpha\beta}\gamma^{\mu\nu}\Gamma^{\underline1}\epsilon,
\end{equation}
where the Lorentz Chern-Simons piece has been absorbed into $\mathcal{G}$ and the last term vanishes since $\Gamma^{\underline{11}}=0$. We can indeed remove the Riemann term via a field redefinition similar to what was done for $\tilde\psi$ but at the cost of a $\nabla^{(+)}F^{(-)}\nabla^{(+)}F^{(-)}$ term that cannot be removed\footnote{Note that we are very restricted in what field redefinitions we may make since $\delta_\epsilon\psi_i^{(0)}$ and $\delta\tilde\lambda^{(0)}$ both vanish, which means we can really only shift by $\psi_\mu$.}.

What this example illustrates is that not every two-derivative consistent truncation necessarily leads to a four-derivative version, even in the case of torus reductions. It is interesting to note that the problem rests with the vector multiplet; it is perfectly consistent to truncate out just the gravitino multiplet (leaving us with a matter-coupled $\mathcal N=2$ supergravity), but the vector multiplet gets non-trivially sourced by the graviton multiplet at the four-derivative level.  Here, the best we can do is to truncate to minimal supergravity coupled to a universal vector multiplet.  An almost identical issue is present if one tries to truncate to pure $\mathcal{N}=1$ or $\mathcal{N}=2$ supergravity in $D=4$, which leads to difficulty recovering the expected result of \cite{Bobev:2020egg,Bobev:2020zov,Bobev:2021oku}.  It should also be noted that at $(\alpha')^3$ order in $D=4$, it is known that the pure $\mathcal N=1$ theory has to be coupled to at least an extra chiral multiplet \cite{Moura:2007ks}\footnote{See also \cite{Moura:2007ac} for a related story in the torus reduction of type II supergravity.}.

One might also consider a similar truncation of the theory in $D=6$. Here, our $\mathcal N=(1,1)$ Lagrangian may be thought of, in bosonic $\mathcal{N}=(1,0)$ language, as a graviton multiplet\footnote{We denote the self-dual part of $b$ as $b^{+}$ and the anti-self-dual part as $b^-$. This is unrelated to the $\pm$ notation associated with $F^{(\pm)}$.} $(g_{\mu\nu},b_{\mu\nu}^+)$ coupled to a gravitino multiplet $A^{(-)\,i}_\mu$ and a tensor multiplet $(b_{\mu\nu}^{-},\varphi)$. As before, it is perfectly consistent to truncate out the gravitino multiplet
\begin{equation}
    F^{(-)\,i}=0,
\end{equation}
at the four-derivative level; in particular, the $F^{(-)\, i}$ equations of motion trivially vanish when we set $F^{(-)\, i}=0$. This yields the truncated Lagrangian
\begin{equation}
    e^{-1}\mathcal L=e^{-2\varphi}\qty[R+4\qty(\partial\varphi)^2-\fft1{12}\tilde h_{\mu\nu\rho}^2+\frac{\alpha'}{8}\qty(R_{\mu\nu\rho\sigma}(\omega_+))^2].
\end{equation}
However, analogous to the vector multiplet in the $D=5$ case, the universal tensor multiplet cannot be truncated away. This can be seen from the dilaton equation of motion, which again contains a $\qty(R_{\mu\nu\rho\sigma}(\omega_+))^2$ that cannot be removed with field redefinitions, or from the dilatino variation, which again has a Riemann term as in the $D=5$ case.  Presumably, the truncation of \cite{Chang:2021tsj} to $\mathcal N=(1,0)$ supergravity coupled to a tensor multiplet and four hypermultiplets will suffer the same fate; while the hypermultiplets can be removed by a further truncation, the tensor multiplet cannot.

While we have focused on four-derivative corrected heterotic supergravity, more generally, the dilaton coupling to higher curvature couplings precludes it from being consistently truncated out of the lower-dimensional theory%
\footnote{It should be emphasized that this is a \textit{perturbative} statement. In principle, one could consider a non-perturbative reduction with compact dimensions of size $\alpha'$, which could provide a loophole.}.
One implication of this is that a top-down approach to higher-derivative holography will necessarily include, at a minimum, the dilaton multiplet in addition to the supergravity multiplet. It would be interesting to see how this fits with the many non-trivial consistency checks of bottom-up holography performed in the absence of the dilaton.

Finally, it would be interesting to consider the case of sphere reductions, such as heterotic supergravity on AdS$_3\times S^3\times T^4$ or $\mathcal M_7\times S^3$.  The resulting gauged supergravity theories are known at the two-derivative level, but we are not sure whether there would be any obstacles to extending the truncation to the four-derivative theory.  We expect that reductions that preserve half-maximal supersymmetry will be allowed, while further truncations may potentially run into obstructions.  Regardless of the outcome, answering such questions will go a long way toward providing a complete understanding of higher-derivative consistent truncations

\section*{Acknowledgements}
We wish to thank N. Bobev for insightful correspondence. This work was supported in part by the U.S. Department of Energy under grant DE-SC0007859.  RJS is supported in part by a Leinweber Graduate Summer Fellowship.

\appendix 
\section{Technical details}
\subsection{Torsionful Riemann tensor}\label{app:riemann}
The torsionful Riemann tensor appears at $\mathcal O(\alpha')$ in the heterotic Lagrangian and equations of motion.  Although we only need its truncated form, here we give the general frame components of the torsionful Riemann tensor.  This is computed from $R(\Omega_+)=\dd \Omega_++\Omega_+\wedge\Omega_+$, where the torsionful connection is given in (\ref{eq:Omega+}).
\begin{align}
    R(\Omega_+)_{\gamma\delta}{}^{\alpha\beta}=&R(\omega_+)_{\gamma\delta}{}^{\alpha\beta}-\fft12F_{\gamma\delta}^i\qty(\qty(g_{ij}+b_{ij})F_{\alpha\beta}^j-G_{\alpha\beta i})\nonumber\\
    &-\fft12\qty(\qty(g_{ij}-b_{ij})F_{\gamma\alpha}^j+G_{\gamma\alpha i})g^{ik}\qty(\qty(g_{kl}-b_{kl})F_{\delta\beta}^l+G_{\delta\beta k}),\nn\\
    R(\Omega_+)_{\gamma d}{}^{\alpha\beta}=&-\fft12e_d^i\qty(\partial_\gamma\qty(g_{ij}+b_{ij})F_{\alpha\beta}^j+\qty(\qty(g_{ij}+b_{ij})\nabla^{(+)}_\gamma F_{\alpha\beta}^j-\nabla^{(+)}_\gamma G_{\alpha\beta i}))\nn\\
    &+\fft14e^l_d\qty(\qty(g_{ij}-b_{ij})F_{\gamma\alpha}^j+G_{\gamma\alpha i})g^{ik}\partial_\beta\qty(g_{kl}-b_{kl})\nn\\
    &-\fft14e^l_d\qty(\qty(g_{ij}-b_{ij})F_{\gamma\beta}^j+G_{\gamma\beta i})g^{ik}\partial_\alpha\qty(g_{kl}-b_{kl}),\nn\\
    R(\Omega_+)_{cd}{}^{\alpha\beta}=&\fft12e^{ic}e^{kd}\qty(\qty(g_{ij}+b_{ij})F_{\alpha\gamma}^j-G_{\alpha\gamma i})\qty(\qty(g_{kl}+b_{kl})F_{\gamma\beta}^l-G_{\gamma\beta k})\nn\\
    &-\fft12e^{jc}e^{ld}\partial_\alpha\qty(g_{ij}-b_{ij})g^{ik}\partial_\beta\qty(g_{kl}-b_{kl}),
\end{align}
along with
\begin{align}
    R(\Omega_+)_{\gamma\delta}{}^{\alpha b}=&-\fft12g^{ik}e^{lb}\partial_\gamma\qty(g_{kl}+b_{kl})\qty(\qty(g_{ij}-b_{ij})F_{\delta\alpha}^j+G_{\delta\alpha i})+e^{ib}\partial_\gamma\qty(g_{ij}-b_{ij})F_{\delta\alpha}^j\nn\\
    &-\fft12e^{ib}\partial_\alpha\qty(g_{ij}-b_{ij})F_{\gamma\delta}^j+e^{ib}\qty(\qty(g_{ij}-b_{ij})\nabla_\gamma^{(+)}F_{\hat\delta\alpha}^j+\nabla^{(+)}_\gamma G_{\hat\delta\alpha i}),\nn\\
    R(\Omega_+)_{\gamma d}{}^{\alpha b}=&\fft14g^{ik}e^{jd}e^{lb}\partial_\gamma\qty(g_{kl}+b_{kl})\partial_\alpha\qty(g_{ij}-b_{ij})-\fft12e^{ib}e^{jd}\nabla_\gamma^{(+)}\partial_\alpha\qty(g_{ij}-b_{ij})\nn\\
    &+\fft14e^{kb}e^{id}\qty(\qty(g_{ij}+b_{ij})F_{\alpha\epsilon}^j-G_{\alpha\epsilon i})\qty(\qty(g_{kl}-b_{kl})F_{\gamma\epsilon}^l+G_{\gamma\epsilon k}),\nn\\
    R(\Omega_+)_{cd}{}^{\alpha b}=&\fft12e^{kb}e^{ic}e^{ld}\qty(\qty(g_{ij}+b_{ij})F_{\alpha\gamma}^j-G_{\alpha\gamma i})\partial_\gamma\qty(g_{kl}-b_{kl}),
\end{align}
and
\begin{align}
    R(\Omega_+)_{\gamma\delta}{}^{ab}=&-\fft12e^{ia}e^{lb}\partial_\gamma\qty(g_{ij}-b_{ij})g^{jk}\partial_\delta\qty(g_{kl}+b_{kl})\nn\\
    &-\fft12e^{ia}e^{kb}\qty(\qty(g_{ij}-b_{ij})F_{\gamma\epsilon}^j+G_{\gamma\epsilon i})\qty(\qty(g_{kl}-b_{kl})F_{\delta\epsilon}^l+G_{\delta\epsilon k}),\nn\\
    R(\Omega_+)_{\gamma d}{}^{ab}=&\fft14e^{ia}e^{kb}e^{ld}\qty(\qty(g_{ij}-b_{ij})F_{\gamma\epsilon}^j+G_{\gamma\epsilon i})\partial_\epsilon\qty(g_{kl}-b_{kl})\nn\\
    &-\fft14e^{ib}e^{ka}e^{ld}\qty(\qty(g_{ij}-b_{ij})F_{\gamma\epsilon}^j+G_{\gamma\epsilon i})\partial_\epsilon\qty(g_{kl}-b_{kl}),\nn\\
    R(\Omega_+)_{cd}{}^{ab}=&-\fft12e^{ia}e^{kb}e^{jc}e^{ld}\partial_\gamma\qty(g_{ij}-b_{ij})\partial_\gamma\qty(g_{kl}-b_{kl}).
\end{align}
In some cases, implicit antisymmetrization with weight one is needed on the two-form indices.  Note that the covariant derivative $\nabla^{(+)}$ is taken with respect to the torsionful connection $\Omega_+$ on frame indices, except in the $R(\Omega_+)_{\gamma\delta}{}^{\alpha b}$ term where the hat on the $\delta$ index indicates that it is corrected using the torsion-free connection $\Omega$.  (The $\alpha$ index is corrected using the $\Omega_+$ connection.)

Since the two-derivative Lagrangian, (\ref{eq:Lags}), and equations of motion, (\ref{eq:eom2d}), are written in terms of a torsion-free connection, it is useful to make note of the standard reduction of the torsion-free Ricci tensor
\begin{align}
    R_{\alpha\beta}(\Omega)&=R_{\alpha\beta}(\omega)-\fft12F_{\alpha\gamma}^ig_{ij}F_{\beta\gamma}^j-\fft12g^{ij}\nabla_\alpha\nabla_\beta g_{ij}+\fft14g^{ij}g^{kl}\partial_\alpha g_{ik}\partial_\beta g_{jl},\nn\\
    R_{\alpha b}(\Omega)&=e^{ib}\left(\fft12\nabla_\gamma(g_{ij}F_{\alpha\gamma}^j)+\fft14g_{il}F_{\alpha\gamma}^lg^{jk}\partial_\gamma g_{jk}\right),\nn\\
    R_{ab}(\Omega)&=e^{ia}e^{jb}\left(\fft14g_{ik}g_{jl}F_{\gamma\delta}^kF_{\gamma\delta}^l-\fft12\Box g_{ij}+\fft12g^{kl}\partial_\gamma g_{ik}\partial_\gamma g_{jl}-\fft14\partial_\gamma g_{ij}g^{kl}\partial_\gamma g_{kl}\right).
\end{align}
In addition, the reduction of $\hat\nabla_M\hat\nabla_N\phi$ yields
\begin{equation}
    \hat\nabla_\alpha\hat\nabla_\beta\phi=\nabla_\alpha\nabla_\beta\phi,\qquad
    \hat\nabla_\alpha\hat\nabla_b\phi=-\fft12e_{ib}F_{\alpha\gamma}^i\partial_\gamma\phi,\qquad
    \hat\nabla_a\hat\nabla_b\phi=\fft12e^i_ae^j_b\partial_\gamma g_{ij}\partial_\gamma\phi.
\end{equation}

\subsection{A note on the torsionful Lorentz Chern-Simons term}\label{sec:CSterm}

The Lorentz Chern-Simons form, (\ref{eq:LCS+}), is computed with the torsionful connection, $\Omega_+$.  If we were to expand it out with $\Omega_+=\Omega+\fft12\mathcal H$, we would get
\begin{equation}
    \omega_{3L}(\Omega_+)=\omega_{3L}(\Omega)+\Tr(R(\Omega)\wedge\mathcal H+\fft14\mathcal H\wedge D\mathcal H+\fft1{12}\mathcal H\wedge\mathcal H\wedge\mathcal H)-\fft12\dd(\Tr\Omega\wedge\mathcal H).
\end{equation}
The final term is not Lorentz covariant but is a total derivative.  Hence it can be removed by a shift of the $B$-field.  In particular, with
\begin{equation}
\tilde H=\dd B-\fft{\alpha'}4\omega_{3L}(\Omega_+),
\end{equation}
we can make the shift
\begin{equation}
    B\to B+\fft{\alpha'}8\Tr(\Omega\wedge\mathcal H),\qquad \omega_{3L}(\Omega_+)\to\omega_{3L}(\Omega_+)+\fft12\dd(\Tr\Omega\wedge\mathcal H),
\end{equation}
to remove the total derivative from the Lorentz Chern-Simons form.

A similar manipulation can be performed in the lower-dimensional theory.  In particular, in (\ref{eq:LCSf}), we have made the shift
\begin{equation}
    \omega_{3L}(\Omega_+)\to\omega_{3L}(\Omega_+)-\fft12\dd(\omega_+^{\alpha\beta}F_{\beta\alpha\,i}^{(-)}\eta^i),
\end{equation}
to remove a total derivative.  Note that removing this mixed component of $\omega_{3L}$ corresponds to a shift of the gauge fields $B_{\mu\,i}$.

\bibliographystyle{JHEP}
\bibliography{cite}
\end{document}